\documentclass[useAMS,usenatbib]{mn2e}

\usepackage{psfig}
\usepackage[dvips]{graphicx}
\usepackage{lscape}

\def\lesssim{\mathrel{\hbox{\rlap{\hbox{\lower4pt\hbox{$\sim$}}}\hbox{$<$}}}}
\def\gtrsim{\mathrel{\hbox{\rlap{\hbox{\lower4pt\hbox{$\sim$}}}\hbox{$>$}}}}
\def\dive{\nabla\cdot}

\title{The structure of cosmic voids in a $\Lambda$CDM Universe}
\author[Ricciardelli et al.]{E. Ricciardelli$^{1}$\thanks{E-mail:
    elena.ricciardelli@uv.es}, V. Quilis $^{1}$, S. Planelles $^{2,3}$\\
$^{1}$Departament d'Astronomia i Astrofisica, Universitat de Valencia, c/ Dr. Moliner 50, E-46100 - Burjassot, Valencia, Spain\\
$^2$ Astronomy Unit, Department of Physics, University of Trieste, via Tiepolo 11, I-34131 Trieste, Italy\\
$^3$ INAF, Osservatorio Astronomico di Trieste, via Tiepolo 11, I-34131 Trieste, Italy\\ 
}

\begin{document}
\date{Accepted ...  Received ...; in original form ...
}
\maketitle
\label{firstpage}
\begin{abstract}

Eulerian cosmological codes  are especially suited  to properly describe the low density regions. This property makes this class of 
codes excellent tools to study the formation and evolution of cosmic voids. Following such ideas,  
we present the results of an Eulerian adaptive mesh refinement (AMR) hydrodynamical and N-body simulation, that contrary to the 
common practice, has been designed to refine the computational grid in the underdense regions of the simulated volume.
Thus, the void regions are better described due to the combined effect
of the Eulerian character of the numerical technique and the use of
high numerical resolution from the AMR approach.
To analyse the outcome of this simulation, 
we have constructed a new void finder optimally suited to find the
hierarchy of voids in AMR simulations.  
The algorithm identifies voids starting from the cells with least density and highest velocity divergence and then expanding the underdense volume until reaching the void walls, defined from the steepness of the density gradient. 
At redshift $z=0$, in a cosmological box of comoving side length $100
\,Mpc/h$,  we identify hundreds of voids with sizes up to  
$\sim 17\,  Mpc/h$ and  typical density contrast of $-0.8$, 
which show a complex morphology and an intricate hierarchy of nested structures. 
The analysis of their mass density profile leads to the conclusion that 
a universal density profile can be applied to voids of any size,
density, morphology and redshift.

\end{abstract}


\section{Introduction}\label{intro}

Large redshift surveys have shown that galaxies are distributed in a weblike configuration around large underdense regions, the so-called voids.
The first large void, with a diameter of 50 Mpc/h, was identified by \citet{Kirshner81} in the region of Bootes. 
Subsequent surveys have confirmed the existence of several voids \citep{Vogeley94, Shectman96} that fill most of the volume of the Universe. 
However, voids have been difficult to study for a long time, as most of the spectroscopic surveys used to focus on  the most clustered regions, where the majority of the galaxies lie.
Only with the coming of large spectroscopic surveys as two-degree field galaxy survey (2dFGRS, \citealt{Colless01})  and Sloan Digital Sky Survey (SDSS, \citealt{York00}) the sample of voids has become large enough to allow statistical studies \citep{HV04, Croton04, Patiri06a, Hoyle12, Pan12}.

In the framework of structure formation, voids are thought to form from negative density perturbations in the initial density field. 
In contrast to the overdense regions that turn around and collapse, these underdense regions evolve from the inside out. Linear theory predicts that as void expands the density in the interior continuously decreases, and matter accumulates at the boundaries, developing sharp edges. 
As a result of such an expansion, voids generate coherent outflows of
matter and galaxies emptying the void volume \citep{Padilla05,
  Ceccarelli06, Patiri12}.  
Theoretically, the void evolution can be predicted by means of the
excursion set formalism \citep{Bond91}. \citet{Sheth04} showed that
the void population can be described by requiring the random walks to
cross two barriers, one referring to  the critical density threshold 
involved with void merging ({\it void-in-void}) and the other to the
disappearance of small voids embedded in large overdensities ({\it void-in-cloud}).

Besides being a striking feature of the cosmic web, voids offer
  an excellent  environment to probe cosmological models.
The velocity outflows from the  voids can be used to constrain
$\Omega_m$ \citep{Dekel94}, while their intrinsic structure and
morphology is sensitive to the equation of state of dark energy
\citep{Park07, Lavaux10, Bos12}. Moreover, 
\citet{Lavaux12} have shown the potential use of the average shape of
a stacked void to infer the cosmological parameters.

Another conspicuous feature is the existence of a void hierarchy, with
small-scale voids embedded in the larger ones \citep{Vandew93,
  Sheth04, AragonCalvo13}. As voids expand, the fate of these substructures is the
gradually fading of the small-scale underdensities in favor of the
large surrounding void, though the void hierarchy never disappears
completely.

In numerical simulations, one of the problems in the identification of voids is the definition
of a void itself, as there is no clear consensus in literature on what
is a genuine void (see \citealt{Colberg08} for a comparison of
different void finders applied to numerical simulations).  In the most simple case, voids can be
defined as regions devoid of galaxies/haloes of a given
luminosity/mass \citep{Gottlober03, Patiri06b},  
in the same way as done in galaxy redshift surveys \citep{Varela12,
  Pan12, Hoyle12}. Void finders can also rely on the dark matter
distribution \citep{Plionis02, Colberg05, Platen07}, identifying voids
as underdense regions, using as a typical density contrast of 
$-0.8$, as suggested by the linear theory \citep{Sheth04}. 
In the majority of the void finders, voids are assumed to be spherical
regions. Although simple arguments show that isolated voids will
evolve into spherical voids \citep{Icke84}, the interaction with the
surrounding environment is expected to let some imprints in the final
structure of the void, with the result that voids become more
elongated as time proceeds \citep{Platen08, Bos12}. 
Therefore, to capture the complexity of voids and of their
substructures, more elaborate algorithms have been developed,  such as the ZOBOV algorithm
\citep{Neyrinck08} and the watershed void finder (WVF,
\citealt{Platen07, Platen08, AragonCalvo13}), based on the tessellation technique.
Here we introduce a new void finder that is optimally suited to find voids in
AMR simulations. It shares with the WVF and ZOBOV procedures 
the capability of reconstructing the complex morphology of voids and of 
its nested hierarchy, although based on simpler arguments. 
In our definition, voids need to strictly satisfy some
physical conditions, such as positive velocity divergence in the
interior and sharp
density gradients at the void boundaries. With these assumptions, we
show how the
algorithm is able to capture also the intricate web of substructures 
within voids.

The aim of the present work is to characterize the void population in a
numerical simulation specifically designed to follow the formation
and evolution of underdense regions. 
 Void expansion is expected to stretch the small and medium-scale
fluctuations located inside them \citep{AragonCalvo13, Neyrinck13},
that would be instead
suppressed in overdense regions.
 As a consequence of such
stretching, these small-scale fluctuations evolve forming a
complex web of tenuous filaments and low-mass haloes. 
We have thus designed a refinement scheme that uses more
resolution in voids, suitable to resolve such an
intricate void substructure. 
With a new void finder 
devised for this task, we analyse the emerging void population, with
a particular focus on the internal structure of voids. 

The structure of the paper is the following. In Section
\ref{simulations} we describe the simulations used in this work. 
In Section \ref{voidfinder} we detail the void finder algorithm and
the tests done to verify its consistency. We present the results on
the void population in Section \ref{results}  
 and conclude in Section \ref{conclusions}.

\section{The simulation}\label{simulations}

The  simulation  described  in  this  paper is  performed  with  the
cosmological  code  MASCLET \citep{quilis04}.   This  code couples  an
Eulerian  approach  based  on  high-resolution  shock  capturing
techniques  for describing  the  gaseous component,  with a  multigrid
particle mesh  N-body scheme for evolving  the collisionless component
(dark matter).  Gas and dark matter are coupled by the gravity solver.
Both  schemes  benefit of  using  an  adaptive  mesh refinement  (AMR)
strategy, which permits to gain spatial and temporal resolution.

The numerical  simulation assumes a spatially  flat $\Lambda
CDM$  cosmology, with  the following  cosmological  parameters: matter
density    parameter,    $\Omega_m=0.25$;    cosmological    constant,
$\Omega_{\Lambda}=\Lambda/{3H_o^2}=0.75$;  baryon  density  parameter,
$\Omega_b=0.045$;  reduced Hubble  constant, $h=H_o/100  km\, s^{-1}\,
Mpc^{-1}=0.73$;  power  spectrum index,  $n_s=1$;  and power  spectrum
normalisation, $\sigma_8=0.8$.

The initial  conditions are  set up at  $z=100$, using a  CDM transfer
function from \citet{EiHu98},  for a cube of comoving  side length $100 \, h^{-1}
\,  Mpc$.   The  computational  domain is  discretised  with  $512^3$
cubical cells.

A special set up is designed in order to produce a simulation specifically
prepared to follow the formation and evolution of voids. Contrary to the 
common practice in cosmological simulations, where the regions with higher 
densities are best described, either by more particles in SPH technique due 
to their Lagrangian nature or by refined grids in AMR Eulerian schemes, 
we refine the coarse grids only in regions with low densities. 
To do so, we evolve the initial conditions until present time using Zel'dovich 
approximation. Those areas at the initial time, $z=100$, which
produce evolved regions with an overdensity, $\rho/\rho_{_B}$, lower than 10 are refined, 
being $\rho$ and $\rho_{_{B}}$ the total density and the background density, 
respectively. 
 This  value  is a
compromise between a correct coverage  of the volume occupied by voids
and the natural changes produced  in the density contrast field during
the evolution.  If we  consider that clusters  follow a  NFW profile 
($\rho/\rho_B  \sim r^{-3}$, \citealt{NFW})  out to  their virial
radius where  the mean overdensity
 reaches values of 200, the use a threshold of 10 would give us
a boundary zone  around clusters of approximately two  or three virial
radii. In such a way, we ensure that our selection is focusing in the
complementary part  of galaxy  clusters. Moreover, voids,  and low
density regions in general, evolve reducing their average density with
time. A value as high as  10, also allows us to catch the early
stages of voids and their boundaries. 
Thus, 
the initial conditions are created on a coarse grid ($256^3$ cells) with a first level of refinement  
(level $l=1$) on those regions which eventually would evolve to low density zones. 
The  dark matter  component in  the initial
refined  regions is  sampled with  dark matter  particles  eight times
lighter than  those used  in regions covered  only by the  coarse grid
(level $l=0$).

The simulation presented  in this paper uses a  maximum of seven levels
($l=7$) of refinement, which gives  a peak physical spatial 
resolution of $\sim
3\, h^{-1}\, kpc$ at $z=0$.
The ratio  between the cell  sizes for a  given level ($l+1$)  and its
parent   level  ($l$)   is,   in  our   AMR  implementation,   $\Delta
x_{l+1}/\Delta  x_{l}=1/2$.  This  is a  compromise value  between the
gain in resolution and possible numerical instabilities.
 For the 
dark  matter we  consider two  particles species,
which correspond to the particles on the coarse grid and the particles
within the first  level of refinement at the  initial conditions.  The
best  mass resolution  is  $\sim 5\times  10^8\, h^{-1}\, M_\odot$,
equivalent to distribute $512^3$ particles in the whole box.
Although we have only used levels up  to l=1 to analyse the voids
  in the present work, the
formation and evolution of these regions depends on the description of
the matter distribution on higher  levels. On such levels, the gravity
is  numerically  better  resolved,  and  therefore,  the  dynamics  is
different.

\begin{figure*}
\includegraphics[width=16 cm]{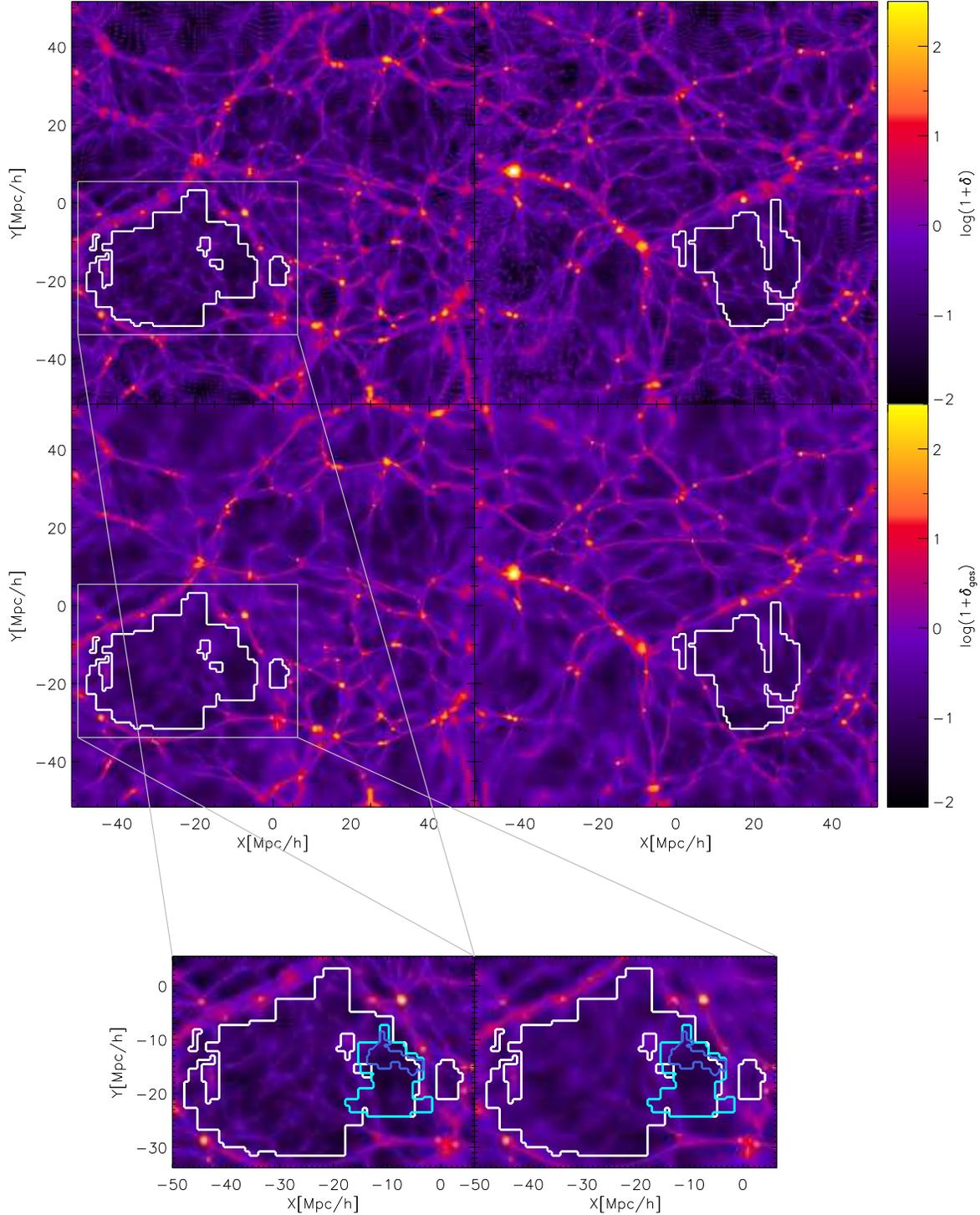}
\caption{The images show the density contrast field in slices of thickness
  $\sim 0.4 \,  Mpc/h$ projected along two coordinate axes of the
  simulated box. 
  In the upper panels the density contrast field includes both the dark matter and the gas
  components. In the lower panels we show the gas density contrast,
  defined as: $(\Omega_m/\Omega_b)(\rho_{gas}/\rho_B)-1$. The white contours
  indicate the  boundaries of one of the largest  voids found by
  the void finder, having equivalent radius of $17  \,  Mpc/h$ and mean
  density contrast of $-0.85$.  The slices are centered at the  cell containing the void center.
  The bottom images show a zoom into the void region for the total
  density contrast (left) and the gas density contrast (right).
  Cyan and blue contours indicate voids in the middle and bottom
  hierarchies, respectively,  contained in the parent void. }
\label{2Dmap}
\end{figure*}

During the evolution, 
we use a mixed strategy  to both describe the voids and the 
denser regions within these voids. In a usual AMR simulation, we would 
refine a cell at the coarse level if its mass is larger than a given free simulation dependent  mass threshold.
Following this procedure, the cells at the newly created level of refinement would be refined when their masses 
would be larger than the same mass threshold, or equivalently, when the total density of the cells increase in 
a factor of eight compared with their parent cells. This procedure would be repeated at all levels of refinement.
However,  in the simulations used in this paper, the coarse grid is refined only in those regions where the 
overdensity is $\rho/\rho_{_B} \leq 10$.  To create the next level ($l=2$) of refinement, we used the standard approach, that is, 
the cells at $l=1$ with overdensities higher that eight times the overdensity threshold, $\rho/\rho_{_B} \geq 80$, were refined.
The following levels of refinement ($l>2$) were created refining the cells with overdensities eight times higher that their parent cells.
This mechanism allows us to follow the formation and evolution of voids by forcing the numerical algorithm to numerically 
resolve them, and at the same time, it makes possible to have a high
numerical resolution to describe the formation of all the structures  within the voids.  
 
 Our simulation includes cooling  and heating processes which take into
account inverse Compton and free-free  cooling, UV 
heating \citep{hama96} at $z \sim 6$, atomic and molecular cooling for a 
primordial gas, and star formation. In order to compute
the abundances  of each species ($H, \, He, \, H^+, \, He^+,\, He^{++}$), 
we  assume that the  gas is optically
thin  and in ionization  equilibrium, but  not in  thermal equilibrium
\citep{katz96,theuns98}.  The tabulated  cooling rates were taken from
\citet{sudo93} and they depend on the local metallicity.
The cooling curve was  truncated below temperatures of $10^4\,K$.  The
cooling and heating  were included in the energy  equation 
\citep[see Eq.~3 in][]{quilis04} as extra source terms.

Although the analysis of the stellar component in voids is not considered in the present work, 
the description of the star  formation and metallicity is introduced in the  MASCLET code  following the
ideas  of \citet{yepes97}  and \citet{springe03}.

\section{The Void finder}\label{voidfinder}

\subsection{The algorithm}\label{algorithm}

Voids in simulations can be searched by looking at the  regions devoid of galaxies or relying on the dark matter (and gas) distribution.
The first approach has the advantage of a straightforward comparison with the observations, that can only rely on visible matter. 
On the other hand, using the galaxy distribution requires a good understanding of the physics of galaxy formation. 
As shown in several works, hydrodynamical simulations still have difficulties in reproducing the observed galaxy mass function \citep{Crain09}, and the use of galaxies as a density tracer can lead to spurious voids. 
Even semi-analytical models that are designed to match the observational data as closely as possible, have shown to fail in reproducing the void distribution observed \citep{Tavasoli12}. Moreover, the sparcity of the galaxy population prevents to  recover the ellipticity evolution in voids \citep{Bos12}.
For all these reasons, we have built our void finder based on the
continuous matter distribution. 
However,  caution must be taken when comparing voids found with this method with observed catalogues of voids, as the two methods are based on different definitions of voids. To keep the case as general as possible we do not assume any a priori shape for the voids. 
This allows us to study also void morphology and its evolution with redshift.
 As for the density field, throughout  the paper and unless
  otherwise specified, we consider the
total density, that is the sum of the dark matter and gas
components \footnote{The continuous density
  field for the dark matter component is calculated
by griddig the dark matter particle distribution onto the  AMR grid, by
means of a triangular-shaped-cloud (TSC, \citealt{Hockney88}) scheme.}
The stellar component is not considered because its
contribution to the total density in voids is negligible.

Our void finder is based on two basic assumptions: (i) voids have
positive velocity divergence in the interior, with the maximum
divergence being located in the void center, as a consequence of
faster expansion of the inner shells with respect to the 
outer regions; (ii) the density at the edges has a sharp increase, hence a steep gradient. 
To be more precise, the algorithm performs the following steps:\\
(i) A grid cell is marked as candidate to be center of a void when the following both conditions are satisfied:
\begin{equation}
\delta<\delta_{thre}
\end{equation}
\begin{equation}
\dive {\bf v}>0
\end{equation}
where $\delta$ is the density contrast in a given cell, 
$\delta=\rho/\rho_B-1$, and $\rho$ stands  for the total density,
dark matter plus gas. 
$\delta_{thre}$ is  a free parameter and   {\bf v} is the peculiar
velocity field of the gas.  \\
(ii) The marked cells are ranked according to the velocity divergence, from the largest to the smallest. Thus, the first void is placed around the cell with the maximum divergence.\\
(iii) The void volume is expanded by adding two cells in each
coordinate direction. \\
(iv) The volume expansion is stopped when at least one of the following conditions applies:
\begin{equation}
\nabla\delta>\nabla\delta_{thre}
\end{equation}
\begin{equation}
\delta>\delta_{max}
\end{equation}
\begin{equation}
\dive {\bf v}<0
\end{equation}
where $\nabla\delta_{thre}$ is the gradient at the void edges and $\delta_{max}$ is the maximum density contrast allowed in a void cell. 
The algorithm is insensitive  to this last parameter, as long as its
value is large compared to the mean density in voids. 
We set it to $\delta_{max}=100$.
 We have also  tested the  algorithm by  using the
       velocity   of    the   dark matter   particles    to compute  the   divergence.
       Given the Lagrangian nature of dark matter  particles, 
       the velocity  divergence can be
       poorly-defined in very  underdense regions when using standard methods.  Since the accurate
       determination  of the  velocity divergence  is crucial  for the
       correct location  of the void  center, we have adopted  the gas
       velocity  approach. Alternatively, Delaunay Tessellation
       Field methods could be used for the velocity field
       reconstruction \citep{Schaap00}.

\begin{figure}
\includegraphics[width=0.95\columnwidth=16 cm]{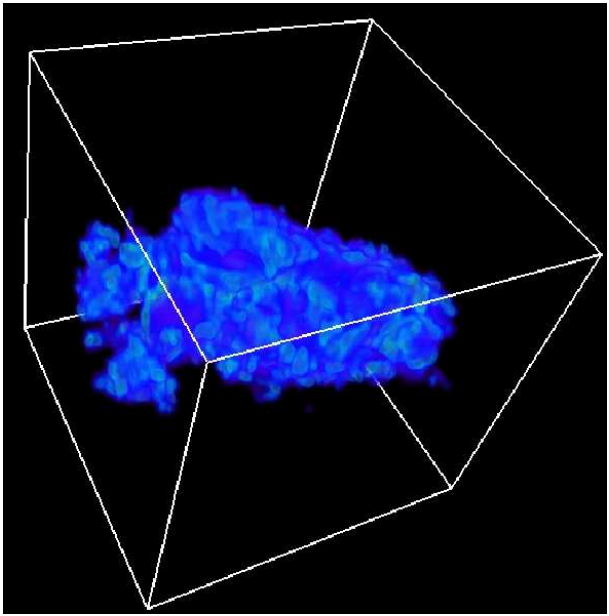}
\caption{3D representation of the same void shown in Figure \ref{2Dmap}.}
\label{void3D}
\end{figure} 

In this way, we find the minimum rectangle parallelepiped contained inside a void. No minimum size is set for these protovoids.
Finally, we allow the protovoids to overlap  with each other. 
When the overlapping volume is smaller than $F_{min}$, the two volumes are considered two separate voids, if it is larger than $F_{max}$, the smallest protovoid is removed from the list, whereas in the case the overlapping volume is within $F_{min}$ and $F_{max}$, we merge the two protovoids, and the new center is given by the volume weighted mean of the two. 
The procedure is then iterated and other protovoids can merge into the same master void. 
At the end of the procedure, we have voids with arbitrary shape, whose
volume is given by the sum  of all grid volume elements 
contained inside the void. The void size is given by the equivalent spherical radius, $R_e$, i.e. the radius of the sphere which have the same volume as the void.
We describe in Section \ref{test} how the free parameters involved in the
procedure, namely
$\delta_{thre}$,  $\nabla\delta_{thre}$, $F_{min}$  and $F_{max}$, have been chosen. 

In order to take into account the hierarchical structure of voids we
have run the algorithm on three grids with different resolution, so as to find
the parent voids and the subvoids in two nested hierarchies. 
We have used the same parameter setting for each hierarchy.
For the parent voids (top level of the void hierarchy)  we
have used the density and velocity fields smoothed on a grid coarser
than the base level of the simulation, having $128^3$ cells (spatial
resolution of $0.8 \,Mpc/h$, i.e. twice the resolution of the base level).
Subvoids, i.e. voids contained in the parent voids, have been
searched using the base level of the simulation ($256^3$ cells, middle
level of the void hierarchy) and the first level of refinement ($512^3$
cells, bottom level of void hierarchy). For each subvoid level (middle 
and bottom hierarchies) the void finder has been applied only to the
volume in  voids at the previous level of hierarchy. Only voids larger
than $3  \,Mpc/h$ are considered for subvoid search. 

As a visual test of the goodness of the algorithm, 
in Figure \ref{2Dmap} we show one of the largest voids identified in the
simulation, overplotted to the density field.  The visual check is very encouraging. 
The algorithm appears able to detect the most
underdense regions in the simulated box and to correctly identify the
void boundaries.
We show both the total density contrast ($\delta$) and the gas density
contrast, defined as:
$\delta_{gas}=(\Omega_m/\Omega_b)(\rho_{gas}/\rho_{B})-1$. 
Although gas dynamics is much more rich and complex than the dark
matter dynamics, in low density regions, the dynamics of both
components is practically identical. For an ideal gas, as the one
considered in our simulation, the pressure is directly proportional to
the density.  In voids,  the density is  very low and,  therefore, the
pressure is extremely  low as well. In this limit,
the equations describing  the dynamics of gas and  dark matter are the
same, and the evolution of both components is extremely
similar. Indeed, as shown in Figure 1, 
the distributions of the gas
density contrast and the total
density contrast in void regions
 appear very similar.

\subsection{Void shape}

 \begin{figure*}
\includegraphics[width=0.69\columnwidth]{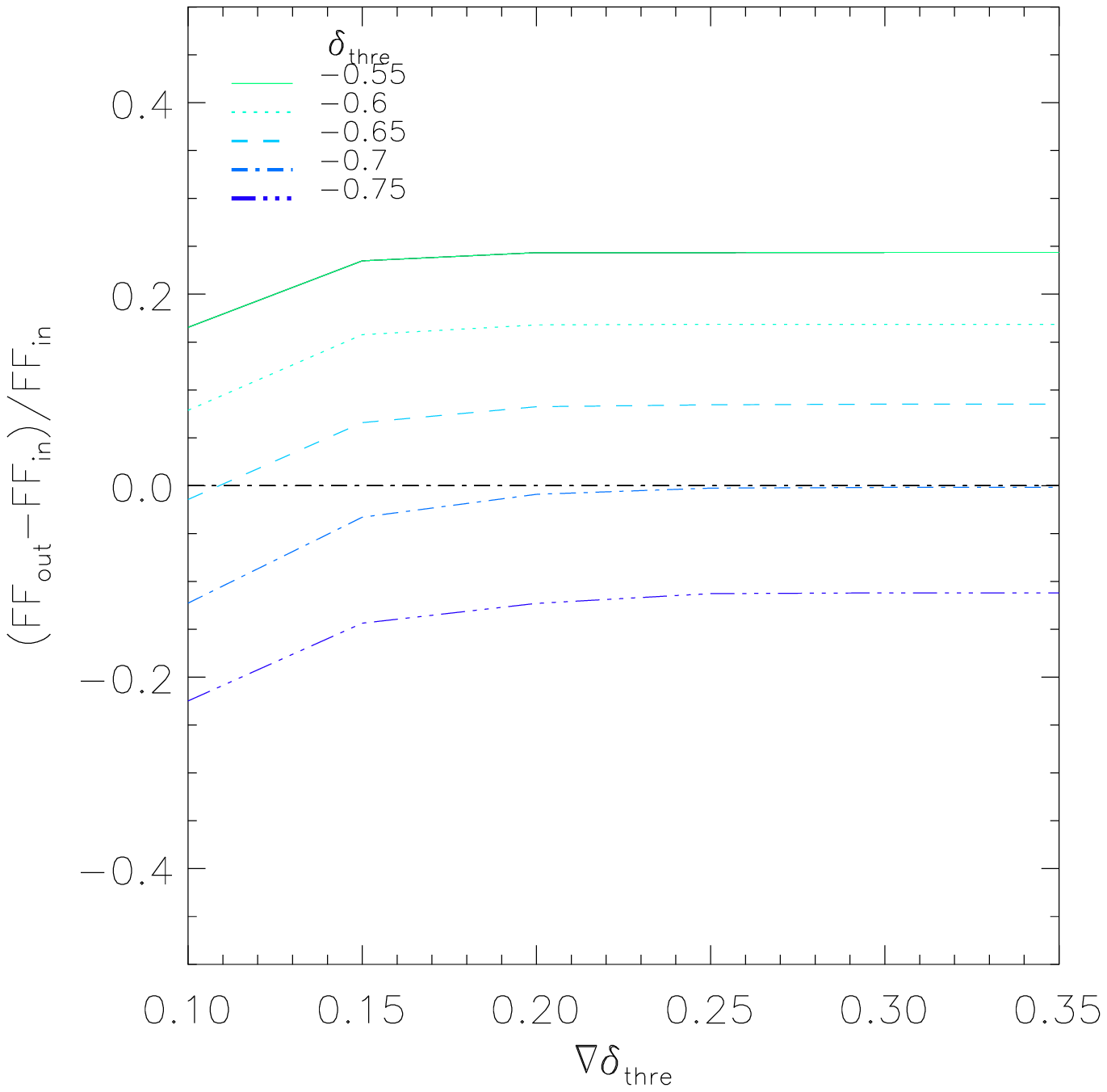}
\includegraphics[width=0.69\columnwidth]{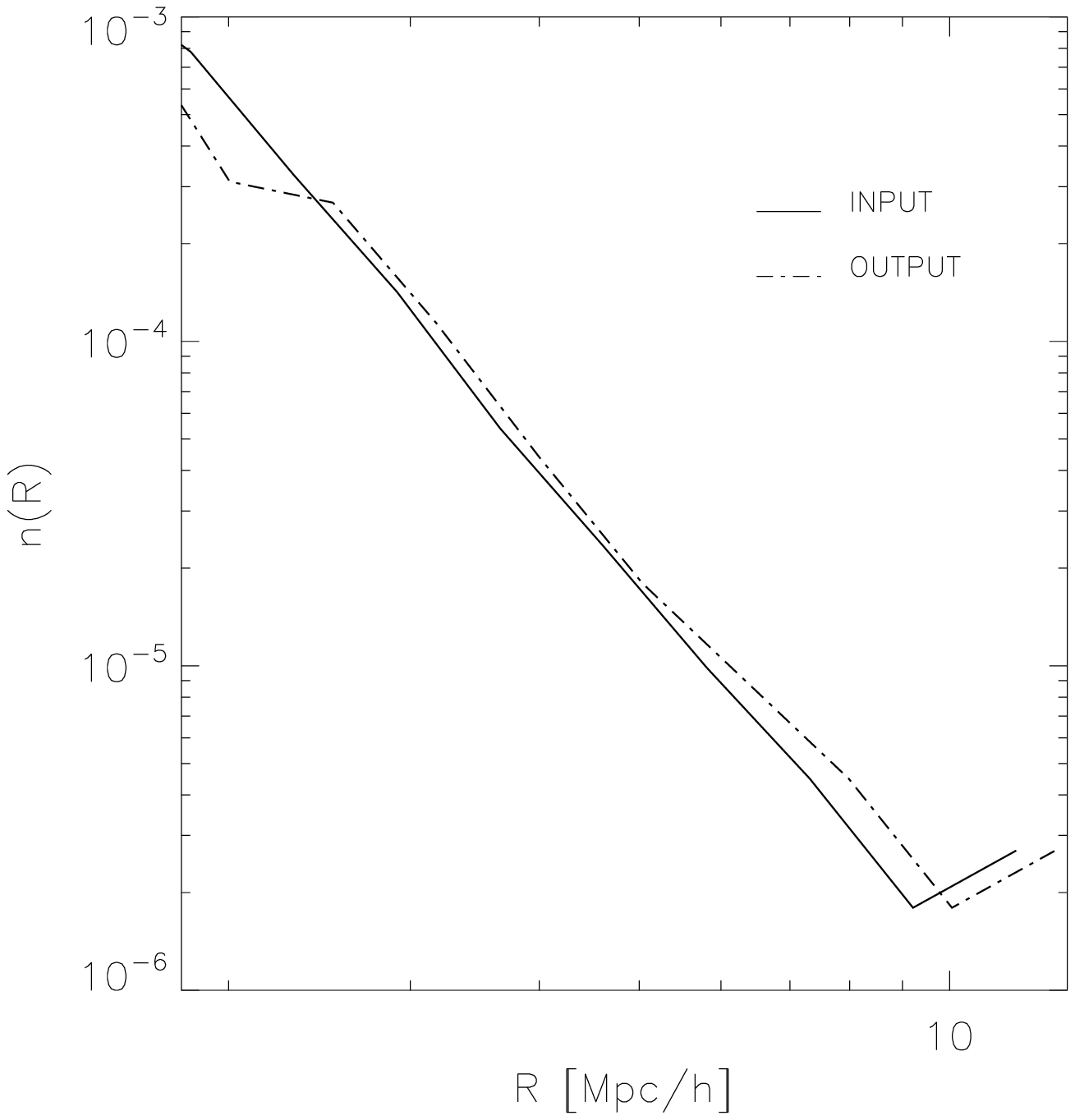}
\includegraphics[width=0.69\columnwidth]{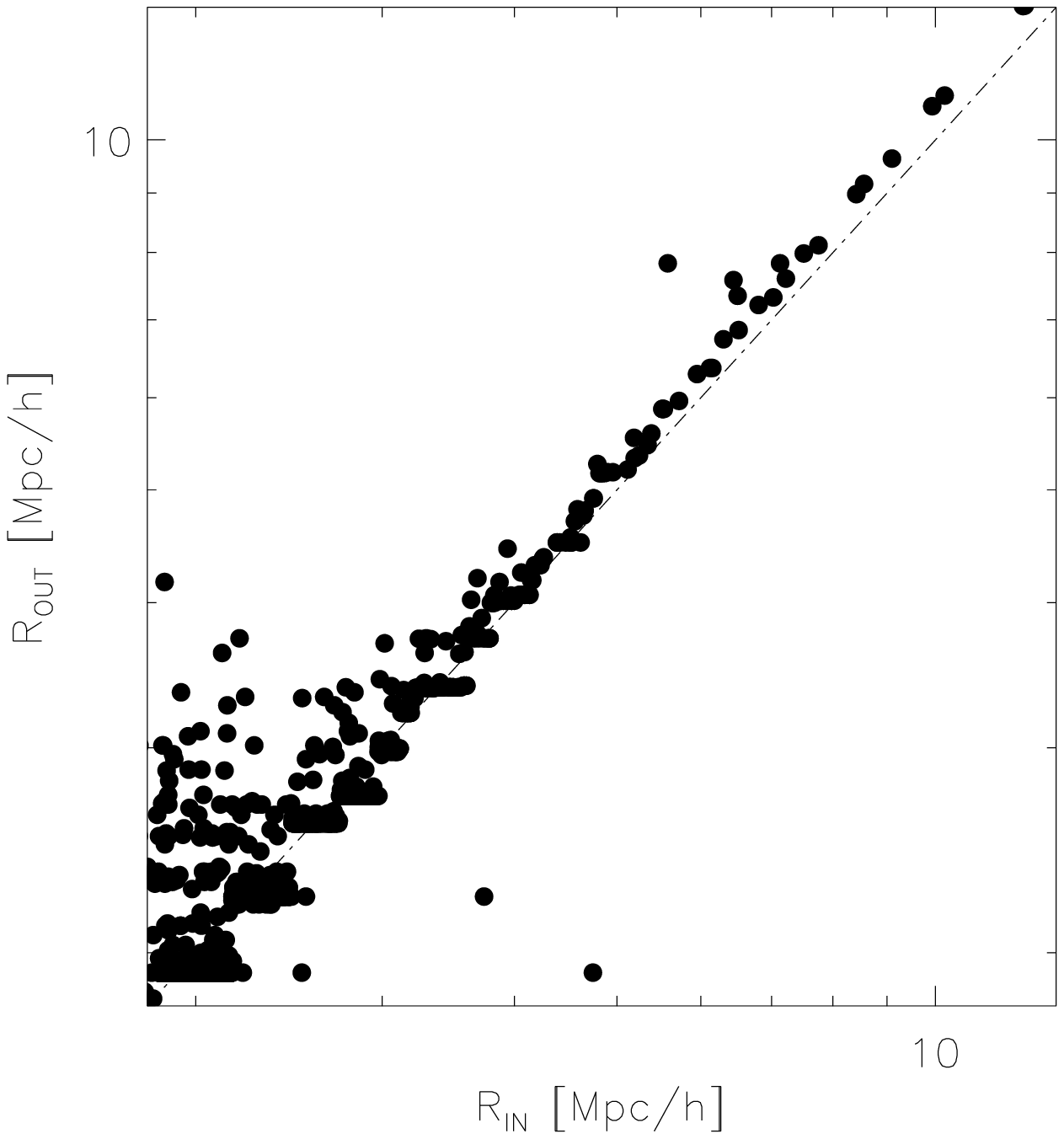}
\caption{Comparison between the input catalogue  of mock voids and that
  inferred from the void finder.  The left-hand panel shows the
  relative difference between the obtained and the input filling
  fractions, as a function of the density gradient for different values
  of the threshold density contrast, encoded by different colours as indicated.
  The set of parameters giving the best match of the FF are used in
  the other two panels: $\delta_{thre}=-0.7$ and
  $\nabla\delta_{thre}=0.25$. The middle panel shows how the inferred
  void density function compares with the input distribution and the
  right-hand panel shows the comparison of void radii. Both, the
  density function and the recovered radii, show an excellent agreement
  with the input values. 
}
\label{mocks}
\end{figure*}

To characterize the shape of cosmic voids we fit them with an
ellipsoid having the same inertia tensor as the void \citep{Shandarin06}:
\begin{equation}
I_{xy}=\frac{1}{N_{cell}}\sum_{i=1}^{N_{cell}}( \delta_{xy} r_i- r_{xi}r_{yi}) \,
\end{equation}
where the summation is over all the void cells $N_{cell}$, $\delta_{xy}$ is the
Kronecker  delta, $r_i$ is the distance of the $i$-th cell to the void
center and $r_{xi}$ and $r_{yi}$ are two components of the radial
vector. To avoid to be biased toward the mass concentrations at the
boundaries of voids, we assign equal weight to each void cell.

The shape of the fitting ellipsoid can be inferred from the
eigenvalues of the inertia tensor: $I_1$, $I_2$, $I_3$.  Thus, the semi-axes
of the fitting ellipsoid can be defined as:
\begin{equation}
a^2=\frac{5}{2}(I_2+I_3-I_1) \,
\end{equation}
\begin{equation}
b^2=\frac{5}{2}(I_3+I_1-I_2) \,
\end{equation}
\begin{equation}
c^2=\frac{5}{2}(I_1+I_2-I_3) \,
\end{equation}
where $a\geq b\geq c$. 
Finally, we can define the void ellipticity as $\epsilon=1-c/a$.

To characterize the goodness of the ellipsoidal fit we use the inverse
porosity, as defined in  \cite{Shandarin06}:
 $IP=V_E/V$, where $V_E$ is the volume of the
ellipsoid fitting the void and $V$ is the actual void volume. 
By definition, the volume of the fitting ellipsoid is always smaller
than the volume of the void, hence: $0 \leq IP \leq
1$.  The fit can be regarded as  good as much IP approches to unity.

As shown by the three-dimensional representation of  a typical void in
Figure \ref{void3D}, 
voids show an irregular
morphology, with level of porosity quite high and are far from being spherical.

\subsection{Dependence on the free parameters }\label{test}

To prove the algorithm and test its dependence on the free parameters, 
we have carried out two tests. On one side, we have performed a set of Montecarlo simulations of
isolated voids with realistic density profiles in order to assess the
dependence of the goodness of the algorithm on $\delta_{thre}$ and
$\nabla\delta_{thre}$. On the other side, we have  proved the
performance of the algorithm when varying $F_{min}$ and $F_{max}$
directly on the simulations.

To assess the dependence on $\delta_{thre}$ and $\nabla\delta_{thre}$, 
we have simulated 3000 mock
voids in
a $100 \, h^{-1}\,  Mpc$ box, 
distributed according to a power-law distribution, having size in the
range: 2-20 $\, h^{-1} Mpc$.  The mock voids have spherical shape and are not allowed to overlap with each other.
Following \citet{Colberg05}, we assume an exponential mass
profile with the following form\footnote{Here the Colberg profile has been
  slightly modified to force  the density enclosed within $R_e$ to be equal to $\rho_e$}:
\begin{equation}\label{fit_colberg}
\frac{\rho(<r)}{\rho_e}=\exp\Big[ \Big(\frac{r}{R_e}\Big)^{1.85}-1\Big]
\end{equation}
where $R_e$ is the void effective radius and $\rho_e$ is the density
enclosed within $R_e$. We set the mean density contrast within voids to 
$-0.8$, as suggested by theoretical arguments \citep{Sheth04}. Outside a radius
equal to $1.2 R_e$ the local density contrast is set to $\delta_{max}$.
The velocity divergence is assumed to linearly decrease from the center to
the borders. Hence, the point of maximum divergence and minimum
density stays in the void center.

We have tested the algorithm by using a density contrast threshold ranging
between $-0.4$ and $-0.8$ and a density gradient in the range 0.1-0.4.
In Figure \ref{mocks} we compare the inferred void
distribution with that of the mocks. 
We use the filling fraction (FF), i.e. the fraction of the 
volume occupied by voids, as the parameter  mapping the capability of the
algorithm in recovering the original distribution. For the range of parameters explored, the recovered FF 
differs from the input one within 20\%. 
The algorithm is sensitive to the value chosen for the density gradient
only for small gradients. For $\nabla\delta_{thre}>0.2$ the goodness of
the void finder only depends on the density contrast threshold adopted.
Density contrast thresholds within $-0.65$ and $-0.7$ provide the best match
to the original FF (within $10\%$).
Therefore, we choose as our fiducial values:
$\nabla\delta_{thre}=0.25$ and $\delta_{thre}=-0.7$.
As shown by the middle  panel of Figure  \ref{mocks} this
choice provides an excellent match to the input distribution. 
The recovered void radii  (righ-hand panel of Figure  \ref{mocks})
also show a remarkable agreement with the input values, although a
slight offset between the two is observed, mainly at small radii. The
reason being that 
the finite size of the cells used in the void reconstruction gives
volumes larger than the original spheres (the cell where the edge
condition applies, see Section \ref{algorithm}, 
 is still considered as a void element).
For small voids, this one-cell approximation translates into larger
errors in volume and, hence, radius.

\begin{figure}
\includegraphics[width=0.92\columnwidth]{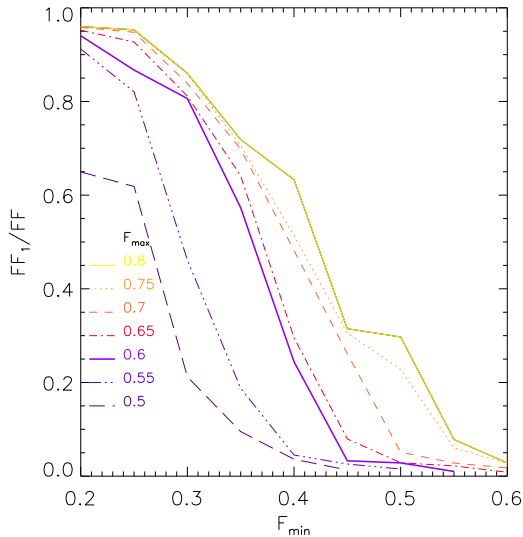}
\caption{Ratio between the filling fraction in the largest void and the
  total filling fraction, for void catalogues obtained for different
  sets of $F_{min}$ and $F_{max}$. The onset of percolation at low
  $F_{min}$ and high $F_{max}$ is clearly visible. }
\label{ff_perc}
\end{figure}

In order to assess the value of  $\nabla\delta_{thre}$ and $\delta_{thre}$,
we have run the void finder on the mock voids, under the assumption that all
the overlapping protovoids can merge to form the final void, without
imposing any criterion for the overlapping volume fraction. 
In this case, the final void is made by all the cells satisfying the
criteria for being  a protovoid and the void volume can be well
reconstructed. 
It is worth to note that such degree of accuracy in the reconstruction of the
void volume can be obtained only in the idealized case of spherical, non
overlapping voids.
In the reality, voids have much more complex structures, and
the use of too generous criteria in the overlapping fractions can lead
to percolating voids. Indeed, being the void walls quite inhomogenous, the void boundaries
are not always well defined and individual voids are 
difficult to identify. 
For instance, voids connected by thin
tunnels would be merged together if too small values of the minimum
overlapping volume were used. Therefore, beside finding the volume
in voids, we require our algorithm to be able to capture the
identity of individual voids.

To find the overlapping fractions providing the best solution for the
real case, instead of using the mocks we have
tested the algorithm directly on the simulations, using different values
for $F_{min}$ and $F_{max}$. In Figure \ref{ff_perc}, we show how the
filling fraction of the largest void ($FF_1$) found in the simulation depends
on these parameters. When using too generous
volume fractions (low $F_{min}$ and high $F_{max}$) for the
overlapping, we see the onset of percolation. In this case
the largest void, encompassing a large number of smaller
voids,  spans a considerable fraction of the
volume in voids, filling almost the entire void volume in the most extreme cases
(see red curve in Figure \ref{ff_perc}).
The onset of percolation occurs quite quickly with $F_{min}$, while
have a smoother dependence on  $F_{max}$. 
For $F_{min}>0.45$ and $F_{max}<0.65$ the
volume occupied  by the largest void is quite stable. Therefore, we
have set our fiducial values to: $F_{min}=0.5$ and $F_{max}=0.6$. 

We have checked that for
small variations around the values adopted for the free parameters, the main results of this work
keep unchanged.

\section{Results}\label{results}

\subsection{Statistics}\label{stat}

\begin{figure*}
\includegraphics[width=0.75\columnwidth]{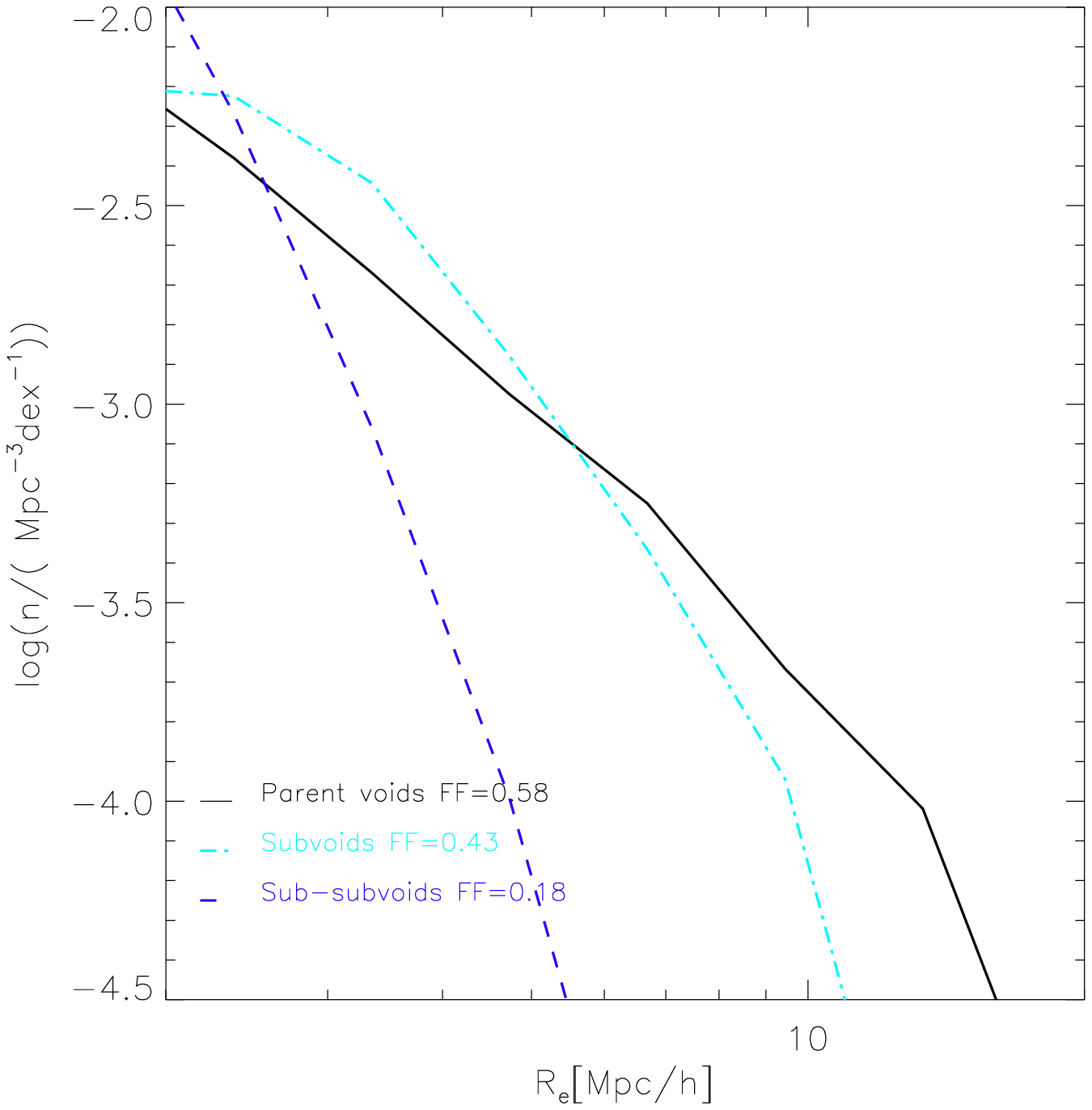}
\includegraphics[width=0.75\columnwidth]{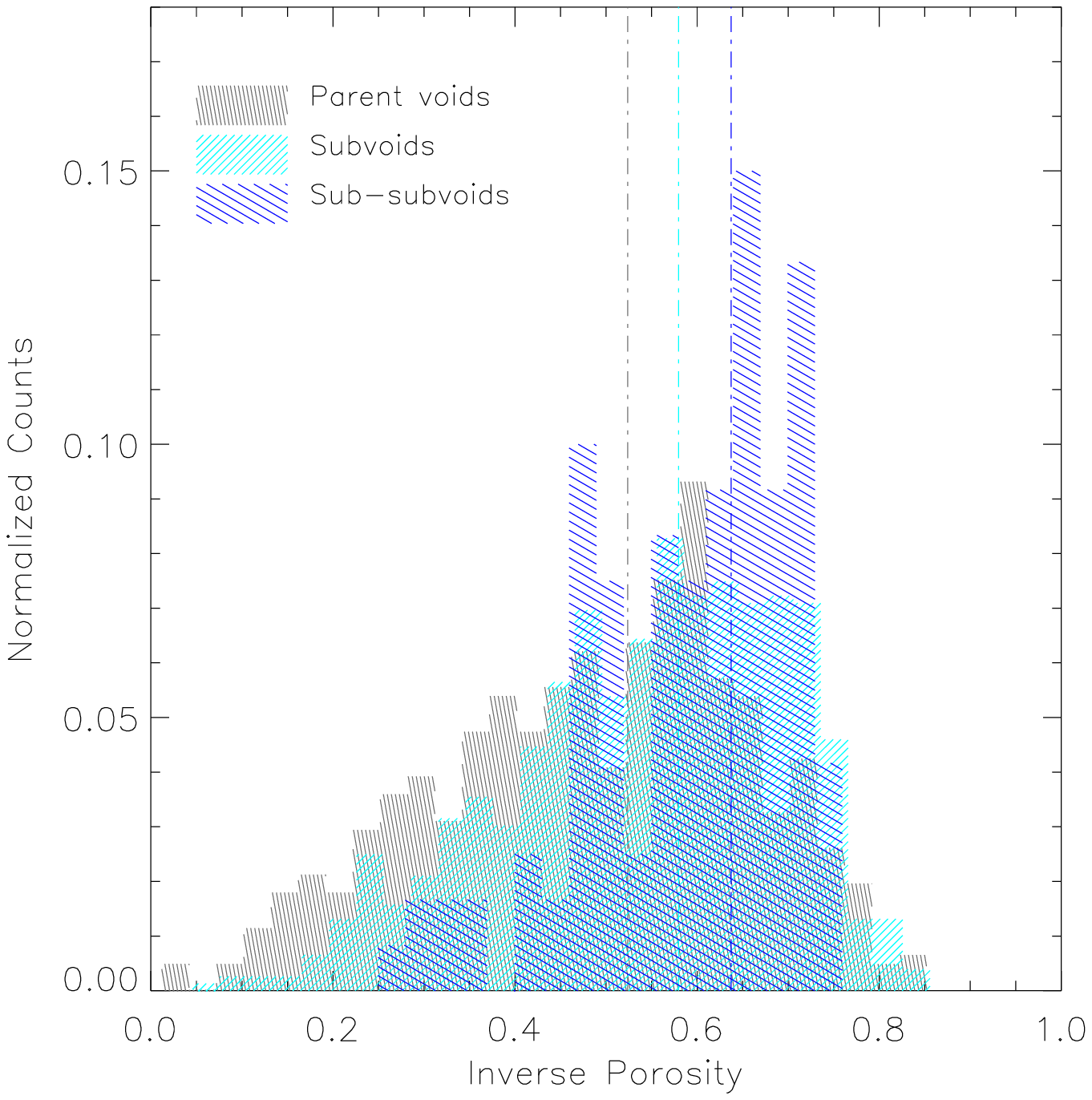}\\
\includegraphics[width=0.75\columnwidth]{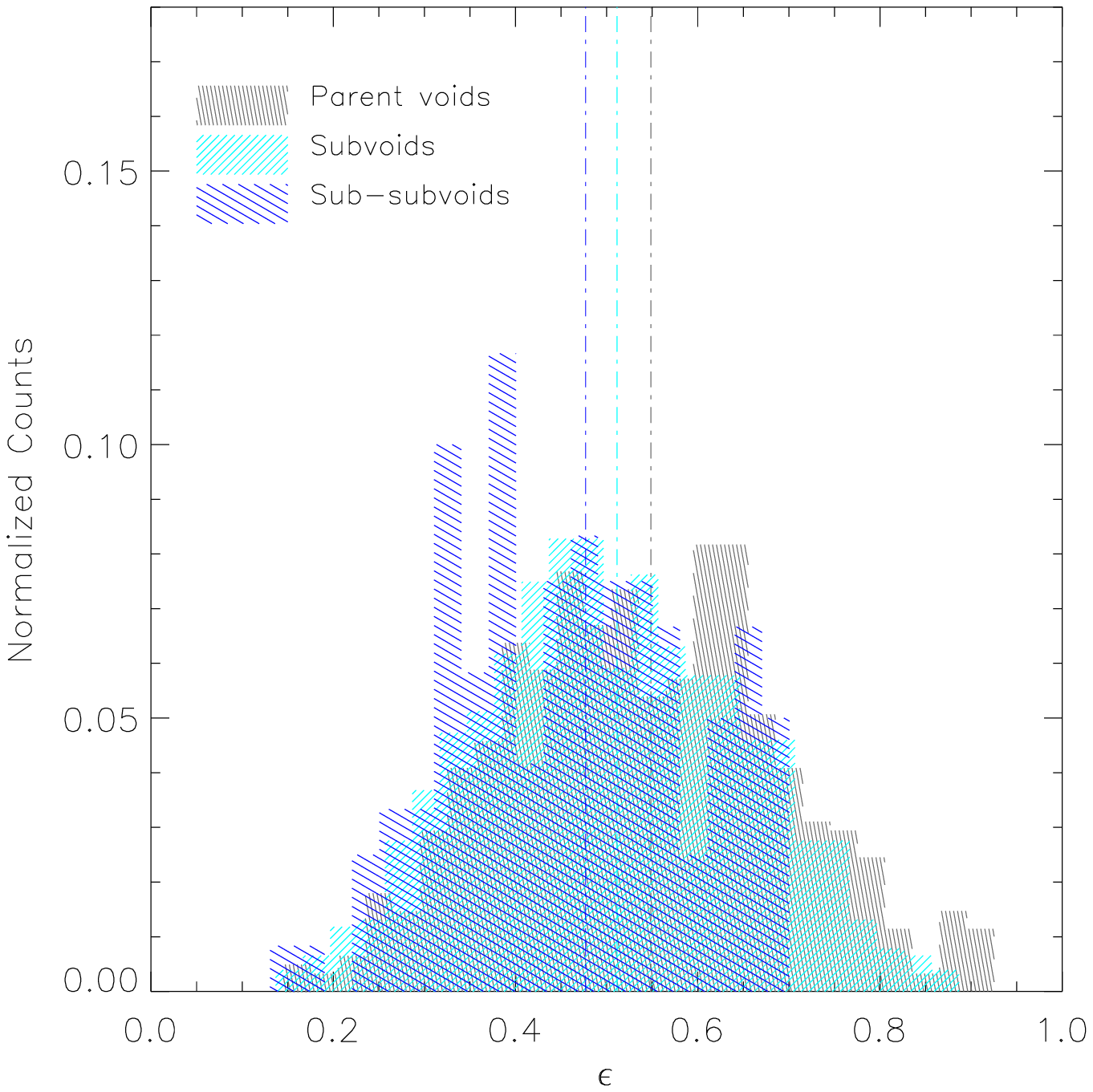}
\includegraphics[width=0.75\columnwidth]{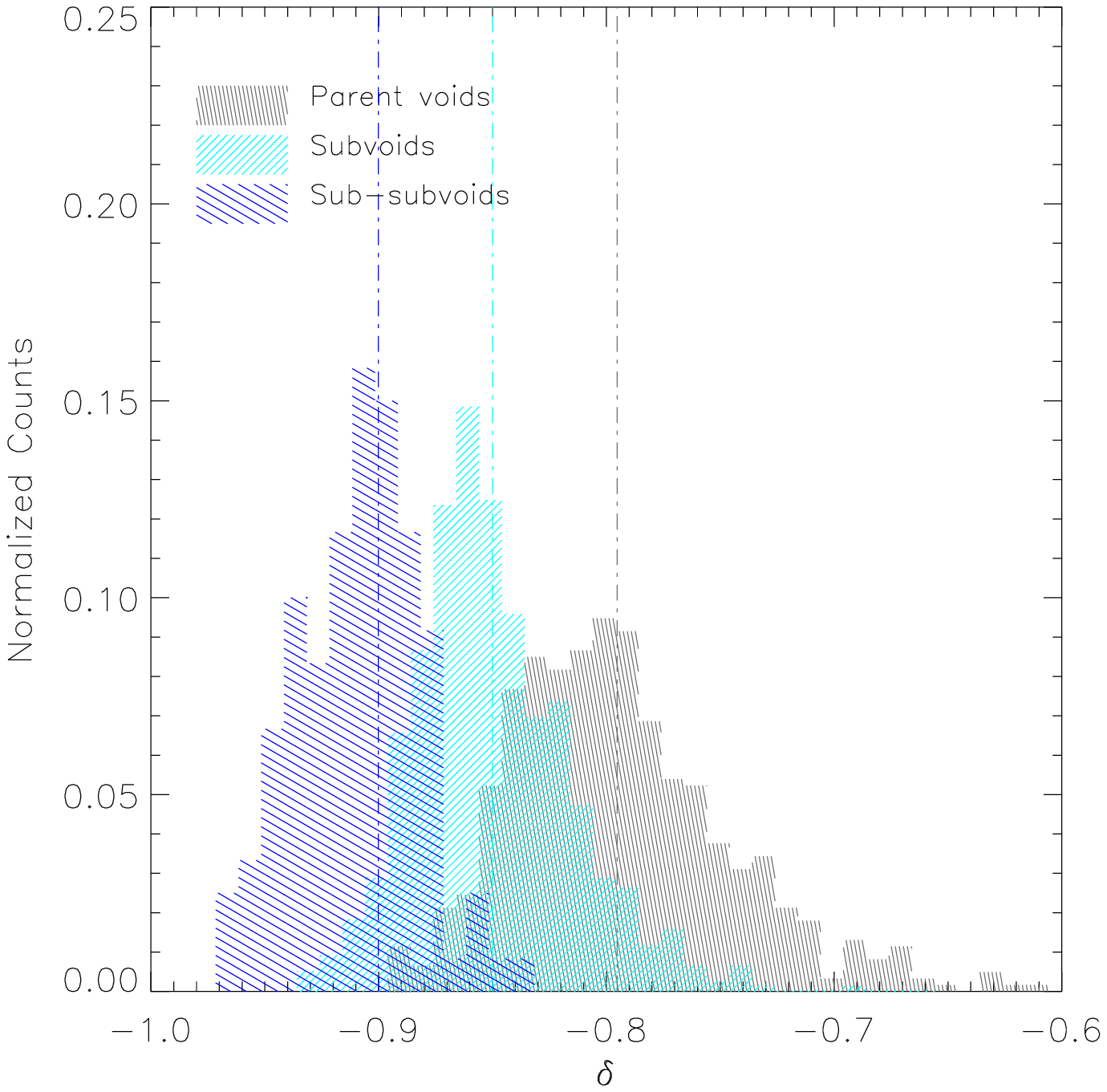}\\
\caption{Basic statistics of the void catalogue for three void hierarchies. The top-left
panel shows the number density of voids of a given effective radius. 
Different colours encode different levels of the hierarchy: black stays
for the parent voids (top level), cyan indicates the subvoids
(middle 
level) and blue stays for
sub-subvoids (bottom level). The filling fraction of voids in each
level of the hierarchy is indicated. 
The other three panels show the distribution of inverse porosity
(top-right), ellipticity (bottom-left) and density contrast 
(bottom-right) for voids at the top  (grey histograms), middle
(cyan) and bottom (blue) level of hierarchy. 
Vertical lines indicate the median value of the distributions. Voids
in lower levels of the hierarchy appear smaller in size, more spherical, less porous
and less dense than their parent voids.
}
\label{statistics}
\end{figure*} 

\begin{figure*}
\includegraphics[width=0.75\columnwidth]{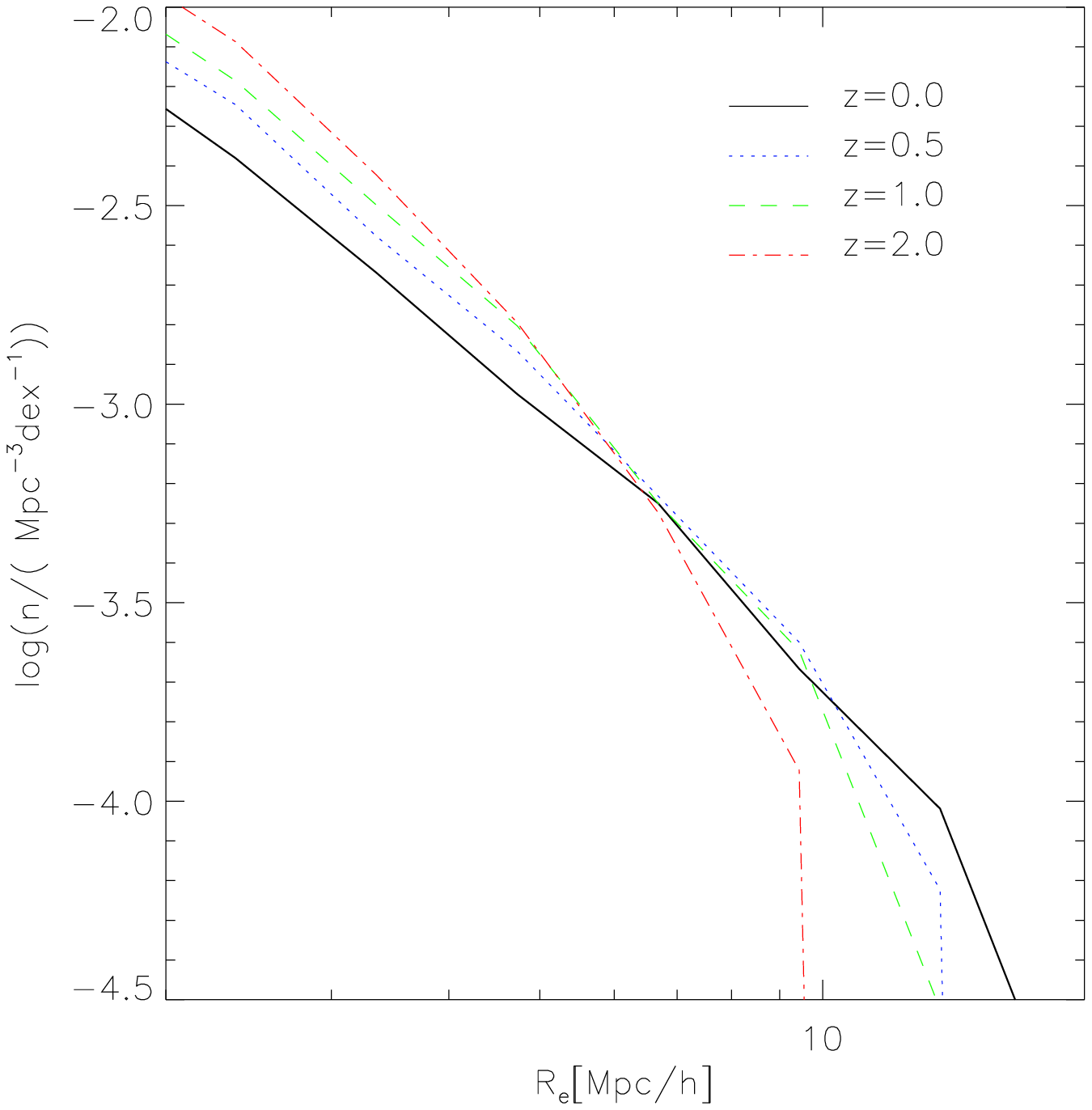}
\includegraphics[width=0.75\columnwidth]{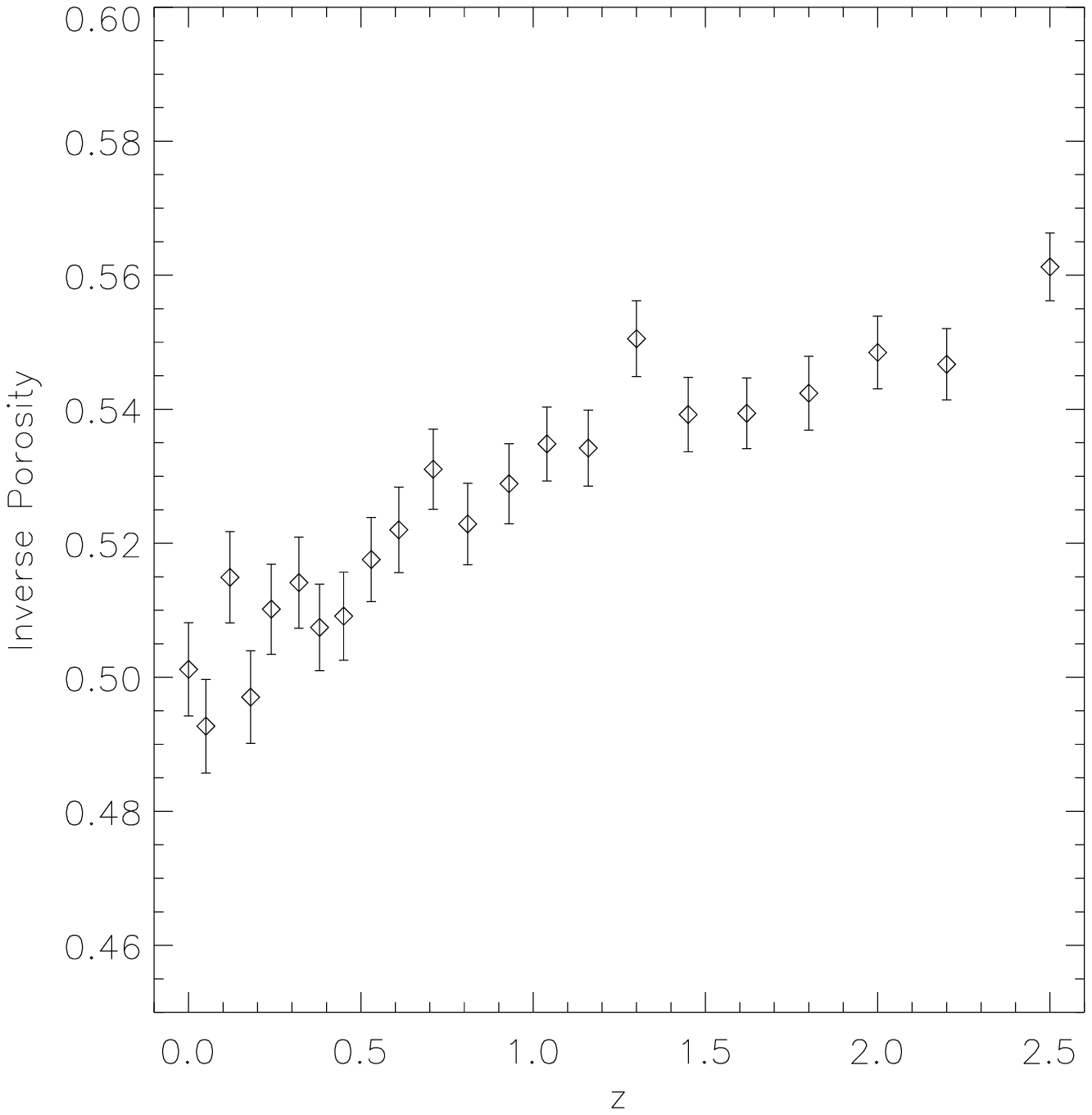}\\
\includegraphics[width=0.75\columnwidth]{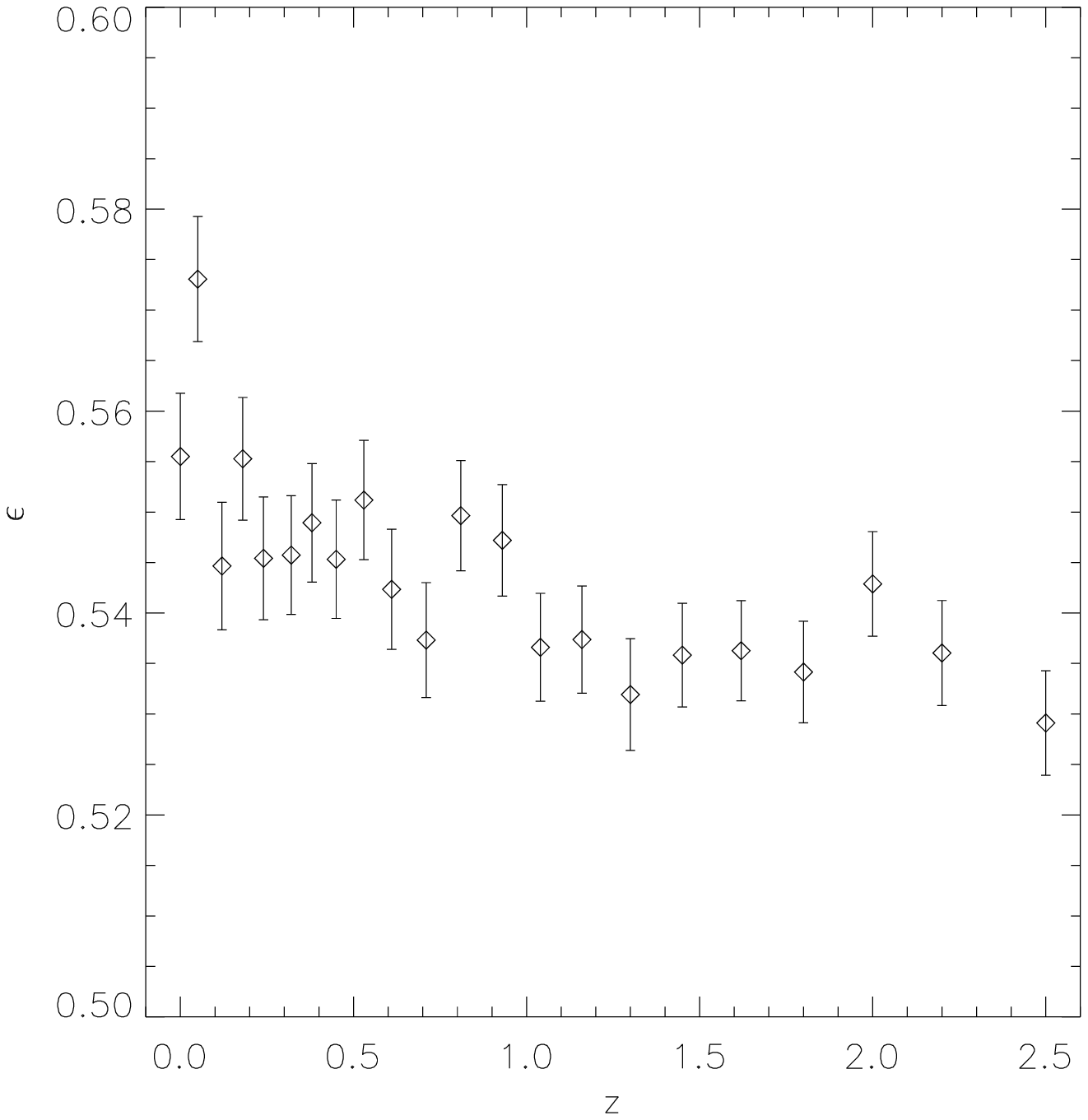}
\includegraphics[width=0.75\columnwidth]{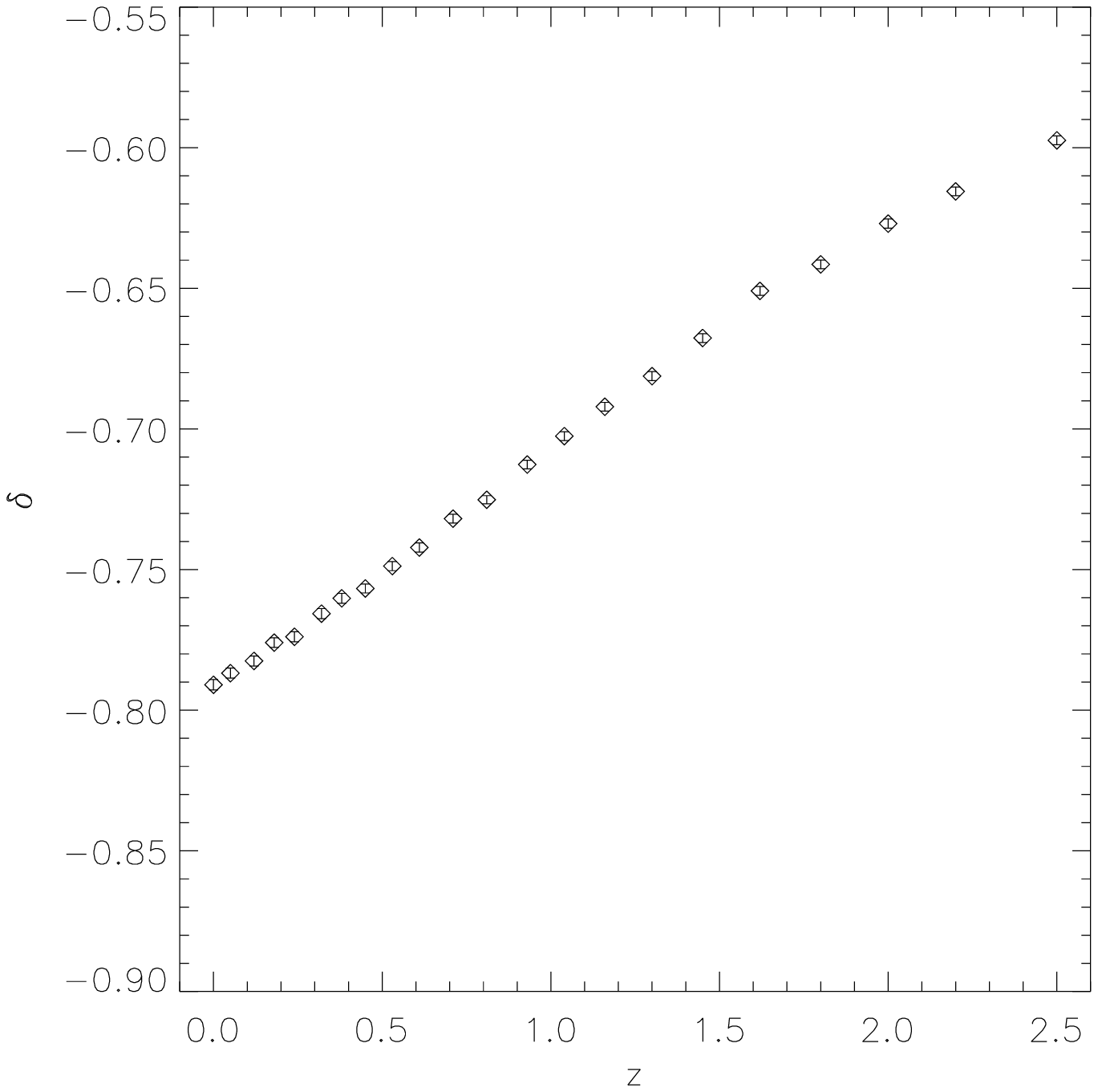}\\
\caption{Redshift evolution of the void population. Only voids at the 
  top level of the hierarchy and with radius larger than $3 \,
  h^{-1}\,  Mpc$  are considered.  The top-left panel
  shows the number density of voids for redshift ranging  from 0 up to 2.5, encoded with
  different colours as indicated. The other panels
  show the redshift evolution of the mean inverse porosity (top-right
  panel), mean ellipticity (bottom-left) and mean density
  contrast 
  (bottom-right). As cosmic time proceeds, voids become larger, more
  elongated and porous and less dense. 
}
\label{statistics_z}
\end{figure*}

In this section we describe the basic statistics of the void
catalogue, obtained with the fiducial values of the free parameters
described in the previous section.
The final catalogue at z$=$0 includes $\sim 600$ voids, with sizes up
to $\sim 17 \, h^{-1}\,  Mpc$ and typical density contrast within
the voids of $-0.8$. The number
density of voids of a given effective radius is shown in the top-left
panel of Figure \ref{statistics}. Voids are distributed according to
a power-law distribution, with a cut-off at large radii. Hence,
smaller voids are much more numerous
than  the larger ones. 
When all the voids are considered the volume in voids is 58\% of the 
simulated volume.
Since we have used a coarse grid resolution to find the parent voids, 
the smallest voids include only few cells, thus their structure can
not be described with accuracy and the shape parameters are less
reliable. Therefore, we cut our void distribution at  
 $3 \,h^{-1}
Mpc$. The filling fraction in voids above this threshold is $\sim
$0.5.
Voids become progressively smaller in the lower hierarchies, with a maximum void radius
of $\sim 10 \, h^{-1}\,  Mpc$ and $\sim 5 \, h^{-1}\,
Mpc$ for voids in the middle and bottom hierarchy, respectively.

In the top-right and bottom-left panels of Figure \ref{statistics} we show the
distribution for the shape parameters: inverse porosity and
ellipticity. 
The sample of parent voids shows a significant level of porosity, with
a median value of inverse porosity of  $\sim$0.55, implying
that the void shape is very irregular and difficult to model. Similar
values have been found by \citet{Shandarin06}  for voids of comparable
size. It is interesting to note that voids in the lower hierarchies
are less porous on average than voids located at the top
of the hierarchy.
The void ellipticity  is shown in the bottom-left panel of Figure
\ref{statistics}.
The median ellipticity of the parent voids is 0.55, indicating that
they are far from being spherical. 
Ellipticity also varies among the void hierarchy, with voids in the
lowest level being the most spherical one ($<\epsilon>=0.45$).
Interestingly, these values are  very similar to
those measured in other works on void shape.
Indeed, \citet{Shandarin06} find a value of 0.55, \citet{Platen08}
give 0.51 and \citet{Bos12} report 0.46. The fact we are able to
reproduce the results of works that are based on completely different void
finders, demonstrates the robustness of the result.

\begin{figure}
\includegraphics[width=1\columnwidth]{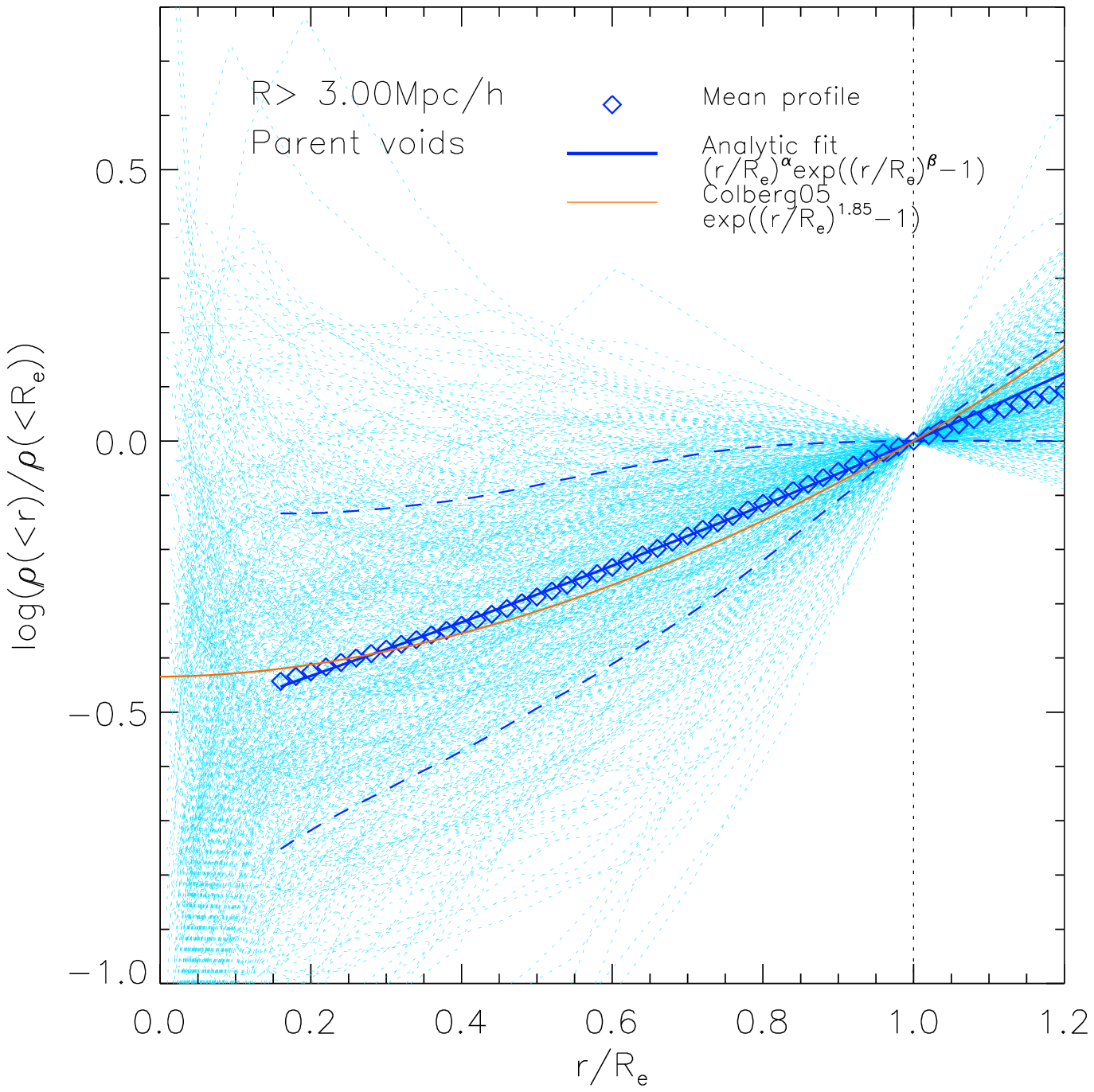}
\caption{Mean density profiles for the parent voids larger than $3 \,h^{-1}\,
  Mpc$. The density profile is expressed in terms of enclosed density
  within a given radius with respect to the enclosed density within
  the void effective radius. Here and in the following figures
    the logarithm is base 10. 
  The dotted cyan lines indicate the individual
  profiles, open symbols show the mean profile, using  a biweight estimator, the thick blue line
  indicates the analytic fit to this profile using Eq. \ref{myfit} and the blue
  dashed lines enclose the 1$\sigma$ confidence interval. For
  comparison we also show the \citet{Colberg05} profile
  as orange line. }
\label{profiles1}
\end{figure} 

\begin{figure}
\includegraphics[width=1\columnwidth]{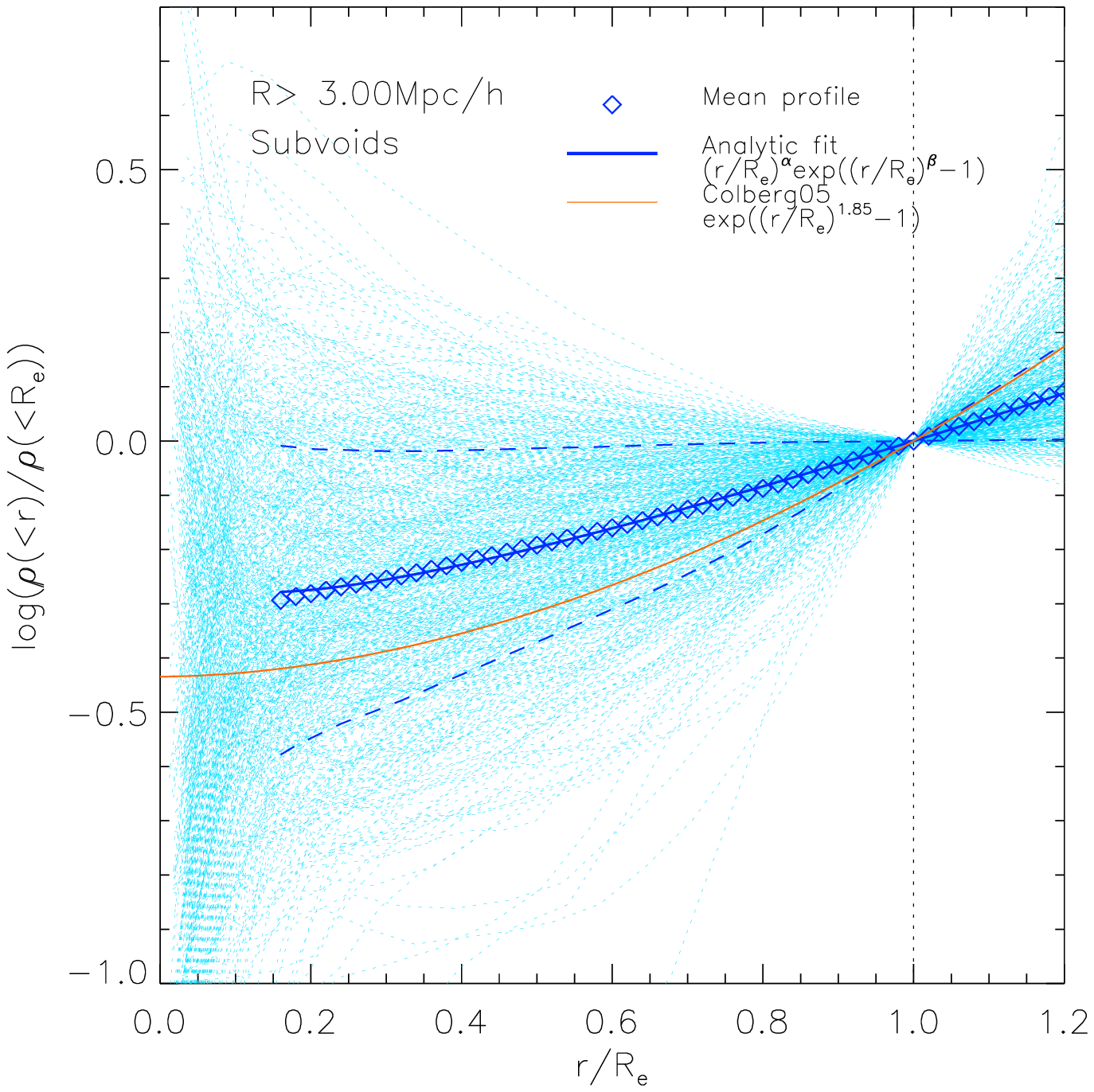}\\
\includegraphics[width=1\columnwidth]{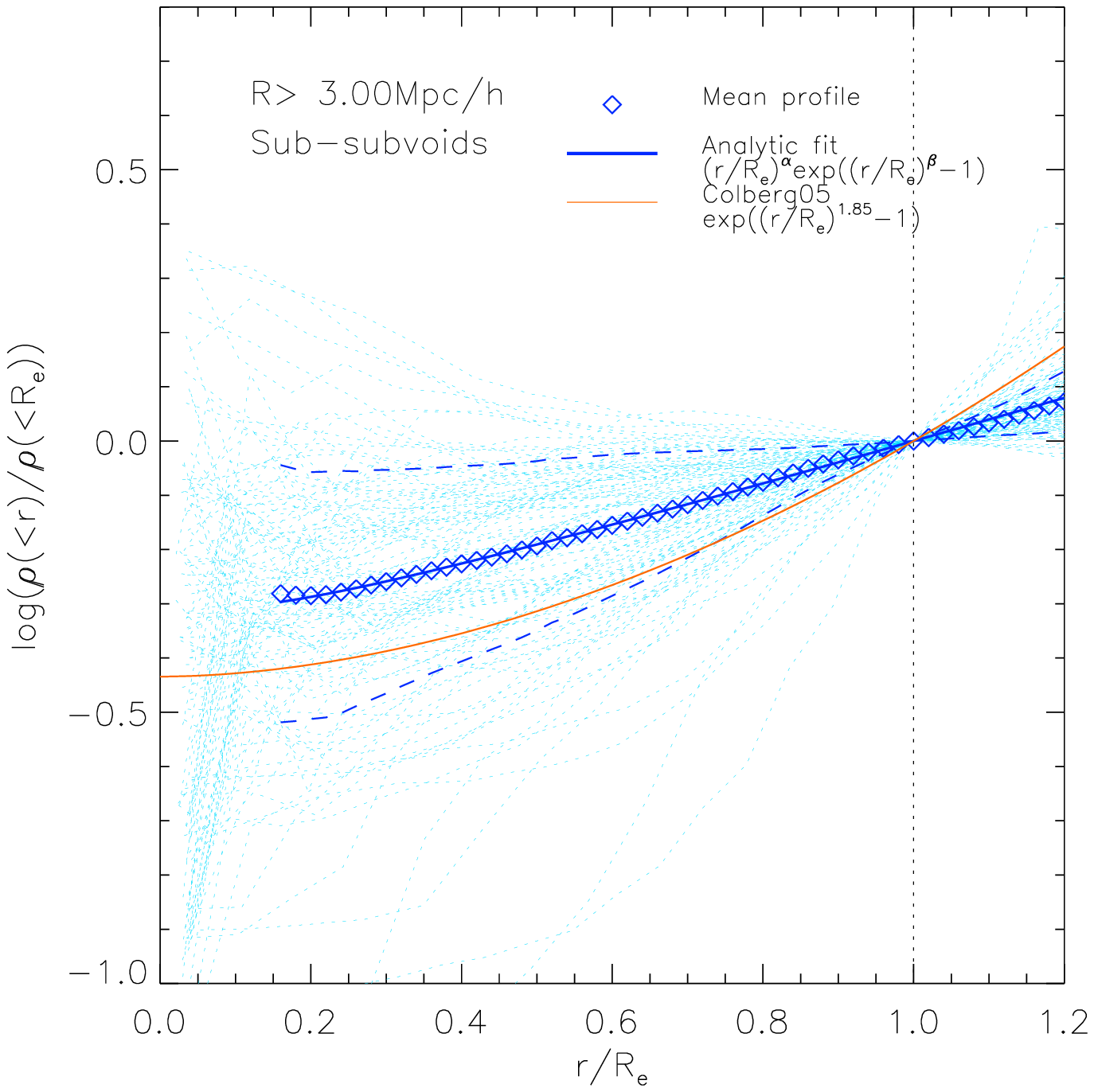}
\caption{The same as Figure \ref{profiles1} for voids in the middle
  (upper panel) and bottom (bottom panel) level of the void hierarchy. }
\label{profiles_lev}
\end{figure} 

The density contrast of voids is shown in the bottom-right panel of
Figure \ref{statistics}.  The typical density contrast of the parent
voids is $-0.8$, as expected since the algorithm have been designed to
target very underdense regions (see Section \ref{test}).  Voids in the lower levels of the hierarchy are much
less dense than their parent voids. The reason  being that voids in
the lower hierarchies only include very thin filaments that do not contribute
much to the total matter content of the void.

The redshift evolution of the void population is shown in Figure
\ref{statistics_z}. 
As expected from theoretical arguments \citep{Sheth04}, we find that voids expand
as time proceeds. As shown in the top-left panel of Figure
\ref{statistics_z},  from high- to low-redshift the abundance of large
voids increases, while at the same time, smaller voids become less
numerous. 
Also the shape parameters display a significant evolution with
redshift, as shown in the top-right panel for the inverse porosity and in the
bottom-left panel for the ellipticity.  
Symbols show the  robust mean using a biweight estimator, 
and its standard error using
all voids larger than $3 \,h^{-1}Mpc$ at any given redshift. 
The biweight estimator \citep{Tukey58} is used in order not to be
affected by the presence of outliers. It
belongs  to the class  of M-estimators,
that defines  the locator by  the minimization of a  function of
the  deviations  of  each  observation  from  the  estimate  of
location   (see  also  \citealt{Beers90}).
The inverse porosity decreases with cosmic time,
hence voids become increasingly more porous at low redshifts.
The ellipticity shows a weak but systematic increasing as redshift
decreases, evolving from $\sim 0.53$ at $z\sim2.5$ up to $\sim
0.55$ at $z\sim0$. In other words, voids become more elongated as time
proceeds.  This appears at odds with 
the naive expectation according to which  voids evolving in isolation  
become more spherical as they expand  \citep{Icke84, Bert85}.
This would be a consequence of the  anisotropic  force field directed outward
that induces a stronger acceleration, and hence a faster expansion, on
the shortest axis direction. However, voids are not
isolated objects, and their evolution is also affected by the large
scale tidal influence, to the extent that voids in realistic
circumstances will never reach sphericity \citep{Lee09, Bos12}. 
Finally, the density contrast evolves considerably from a typical value
of $-0.6$ at $z\sim2.5$ down to $-0.8$ at $z\sim0$.

\subsection{Density Profiles}

In this section we present the density profiles for our sample of
voids larger than $3 \,h^{-1}\,
  Mpc$.  
Because of the particular AMR implementation of this simulation, most of
the void centers are  located at the first level of
refinement ($l=1$, with spatial resolution of $\sim 0.2 \,h^{-1}\,
Mpc$). Hence, we compute the profiles of individual voids using the
density in the  AMR grid at this level.
The density profiles are shown in Figure \ref{profiles1} for voids in
the top level of the hierarchy and in Figure \ref{profiles_lev} for
voids in the middle and bottom levels. 
The density profile is expressed in terms of enclosed density
  within a given radius with respect to the enclosed density within
  the void effective radius. The individual profiles of the voids in
our sample are shown as cyan lines. 
As shown in the previous section, typical voids are not spherical in
shape. Hence, the radial apertures at large distances ($\gtrsim R_e$)
can include cells not belonging to the voids. 
Moreover, the void
center may in some cases fall quite close to the boundaries and lead 
to bumps in the density at small radii. 
This can be easily seen
in some of the individual profiles shown in Figure \ref{profiles1},
where an excess of mass density is seen in the central regions (see
upper curves). Therefore, our results should be interpreted as the 
density profiles of the regions surrounding the voids, more than
the profiles of the voids themselves.

To stack the individual profiles, we rescale the radial
profile of each void to its effective radius and compute a biweight
estimator.
To avoid edge effects we have only considered voids
 whose centers are more than one effective
radius from the box boundaries (95\% of the voids satisfy this condition).
The resulting mean profile with the 1$\sigma$ confidence
interval is shown in Figure \ref{profiles1}. 
As a comparison, we plot also the  fit of \citet{Colberg05},
measured for voids of similar sizes and given by equation (\ref{fit_colberg}).
Our stacked profile agrees remarkably well with the exponential fit
proposed by \citet{Colberg05}, although some deviations from the
Colberg profile exist. 
We note in passing that by adopting the
same minimum void radius as in the Colberg profile ($5 \,h^{-1}\,  
Mpc$) our stacked profile does not change significantly. 
We have found that a better description of the void density profile is
given by the following expression:
\begin{equation}\label{myfit}
\frac{\rho(<r)}{\rho_e}=\Big(\frac{r}{R_e}\Big)^{\alpha}\exp\Big[ \Big(\frac{r}{R_e}\Big)^{\beta}-1\Big]
\end{equation}
where $\alpha$ and $\beta$ are the parameters to be obtained from the
fit. In Figures 
\ref{profiles1}-\ref{profiles_lev} the blue thick lines show the best fits. 
The best fit values are: $\alpha=0.07$ and $\beta=1.32$ (top level
of the hierarchy, Figure \ref{profiles1}), $\alpha=-0.09$ and $\beta=0.99$
(middle level, top panel of Figure \ref{profiles_lev}), $\alpha=-0.12$ and $\beta=1.11$
(bottom level, bottom panel of Figure \ref{profiles_lev}). Subvoids thus show much flatter profiles
than their parent voids and no significant difference is observed
beween the density profile of voids in the middle and bottom level of
the hierarchy.  
The steepness of the parent void profiles can be ascribed
to the high density of the edges. In the case of subvoids, the
steepness of the edges is limited by the condition to be contained in
voids, thus high density walls are not present.

\begin{figure*}
\includegraphics[width=1.8\columnwidth]{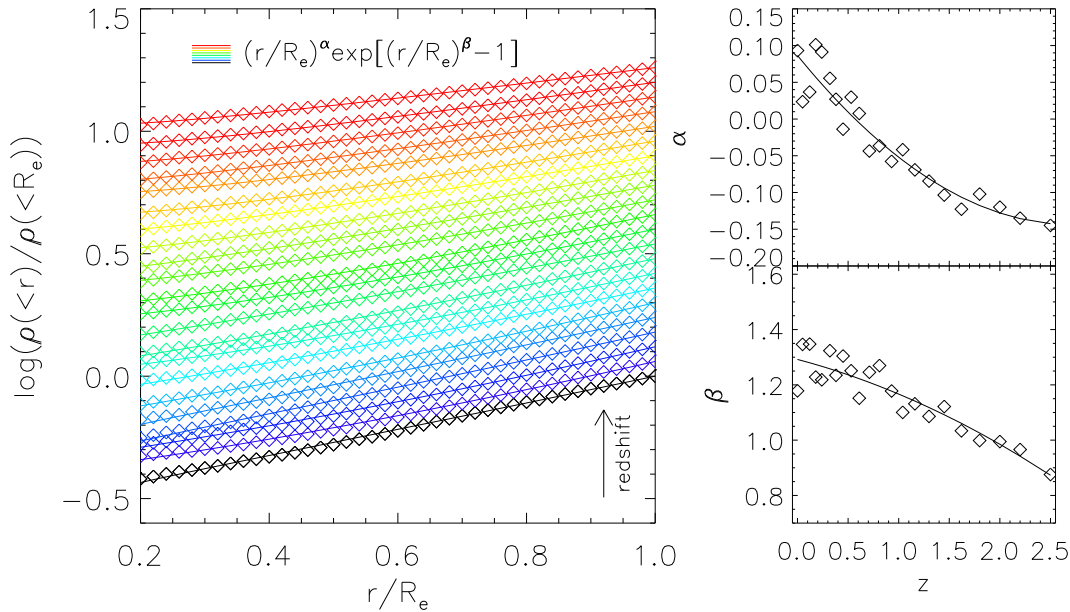}
\caption{Left-hand panel: mean density profiles for voids larger than $3 \,h^{-1}\,
  Mpc$ at different cosmic times, with redshift  increasing from the lower to the upper part of the
  diagram. The values of redshift can be inferred from the
    right-hand panels. The profiles at redshifts greater than 0 have
  been artificially shifted upward for the sake of clearness. Symbols indicate the
  void density profiles and the solid lines are fits to the density profiles
  using equation \ref{myfit}. 
  Right-hand panels: redshift evolution of the free
  parameters involved in the fit, namely $\alpha\,$ and $\beta\,$. The solid
  lines indicate a polynomial fit (see eq. \ref{alphaz} and \ref{betaz}). }
\label{profilesz}
\end{figure*}

In Figure \ref{profilesz}, we show the redshift dependence
of the void radial profiles from z=2.5 down to the present time.
At all redshifts the void sample includes only voids larger than $3 \,h^{-1}\, Mpc$.
The void profiles become progressively steeper as time proceeds,
suggesting that the interior of voids is emptying. 
This is in agreement with the theoretical expectations of
\citet{Sheth04}, showing the evolution of the void density profile towards a reverse top-hat shape, as a consequence
of the continuos evacuation of matter from the void center.
To parametrize this behavior, we show the best fit profile for
each redshift in the left-hand panel of Figure \ref{profilesz}  
and the redshift dependence of the best fit parameters in the right-hand
panels. A polynomial fit to the redshift dependence of $\alpha$ and
$\beta$ is also given by:
\begin{equation}\label{alphaz}
\alpha=0.08-0.16z+0.03z^2
\end{equation}
\begin{equation}\label{betaz}
\beta=1.29-0.09z-0.03z^2
\end{equation}
Although the evolution in the parameters is clearly visible from
Figure \ref{profilesz}, the low
redshift values present a significant scatter. 
We ascribe this high scatter to two reasons. 
On one hand,  at low redshift the void statistics is quite poor, as
the number of small voids decreases. 
On the other hand, the low redshift voids are more porous than their
high redshift counterparts, as shown in Figure \ref{statistics_z},
that would imply that the  profiles at these redshifts are more noisy and the
resulting scatter in the best fit parameters higher.

\begin{figure*}
\includegraphics[width=1.8\columnwidth]{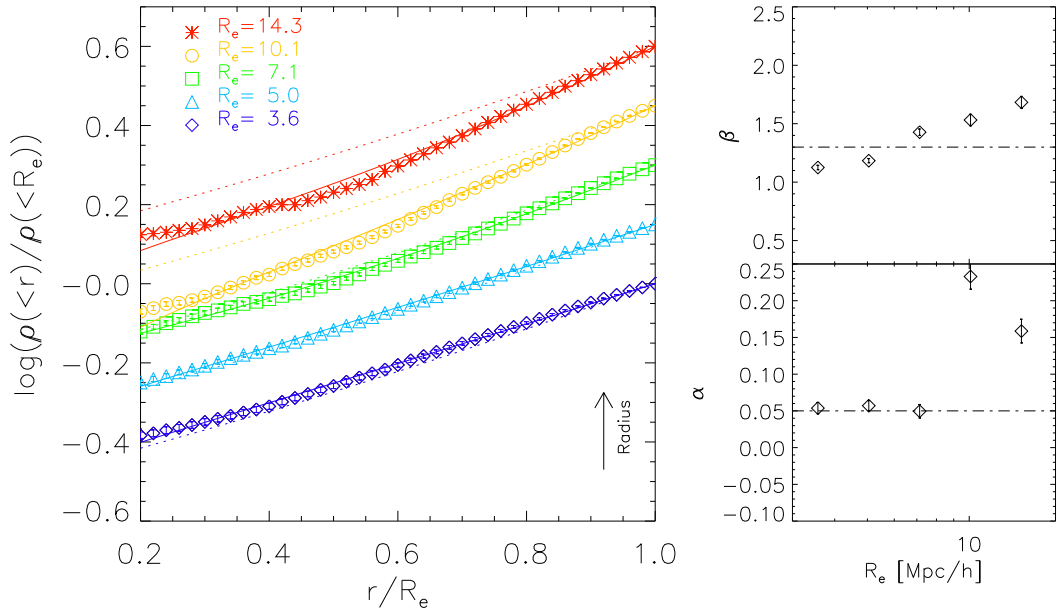}
\caption{Left-hand panel: mean density profiles for voids stacked by
  effective radius, with radius increasing from the lower to the upper part of the
  diagram, as indicated. Equally spaced logarithmic
  bins have been used. 
  All the profiles, except that of  the subsample at the smallest radius, have
  been shifted upward for the sake of clearness. To enhance the statistics we have used all voids larger
  than $3 \,h^{-1}\, Mpc$ in the last five snapshots of the simulation. Symbols with errorbars indicate the robust mean 
  at each radial bin with its standard error. 
  The solid lines indicate the best analytic fit for each
  subsample, whereas the dotted line is the best fit for the global
  population,  reported for comparison. 
  Right-hand panels:  dependence of the free parameters involved in
  the fit ($\alpha\,$ and $\beta\,$) on the void effective
  radius. Errorbars indicate the formal error of the fits. The
  dot-dashed line indicates the values of $\alpha\,$ and $\beta\,$
  for the global population, used for the analytic fit shown as dotted
line in the left-hand panel. }
\label{profilesr}
\end{figure*}

\begin{figure*}
\includegraphics[width=1.8\columnwidth]{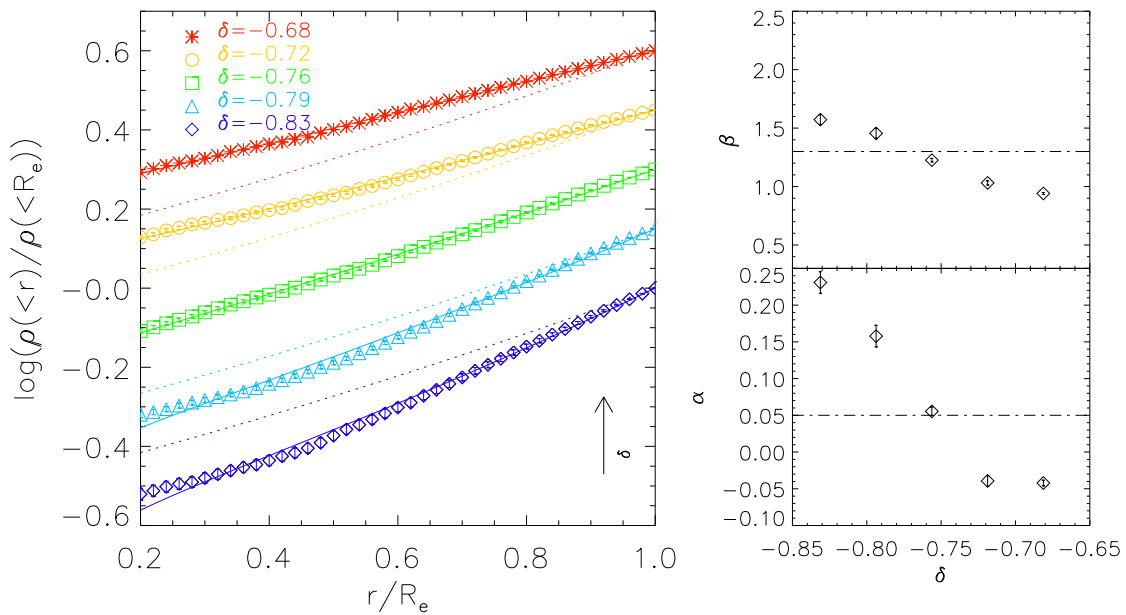}
\caption{The same as Figure \ref{profilesr} for voids stacked 
  according to their mean density contrast. Equally spaced bins in
  density contrast have been used.}
\label{profilesd}
\end{figure*}

\begin{figure*}
\includegraphics[width=1.8\columnwidth]{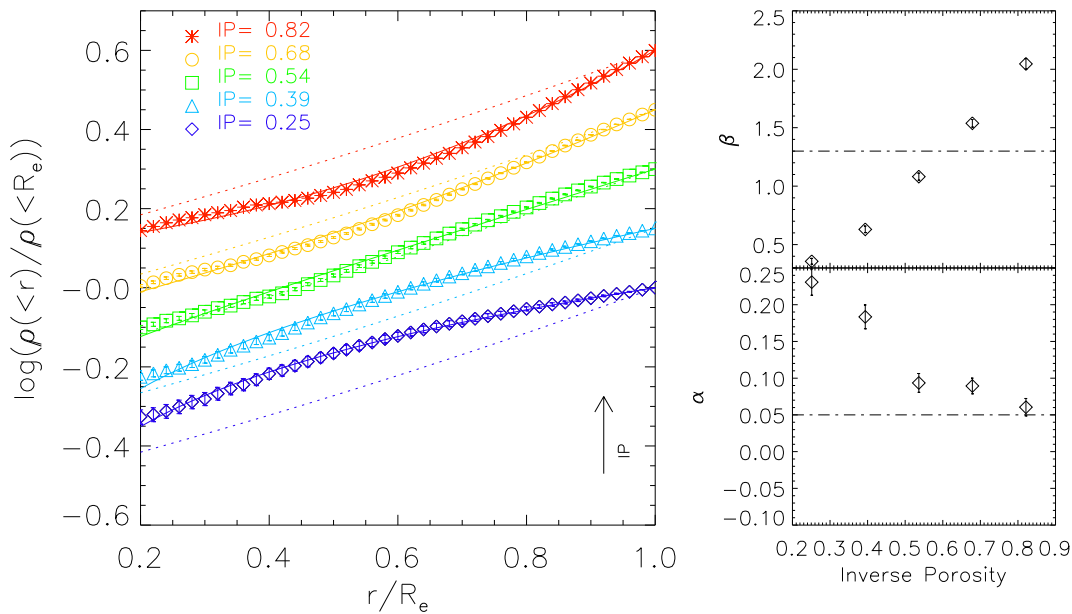}
\caption{The same as Figure \ref{profilesr} for voids stacked 
  according to their inverse porosity. Equally spaced bins in
  inverse porosity have been used.}
\label{profilesi}
\end{figure*}

\begin{figure*}
\includegraphics[width=1.8\columnwidth]{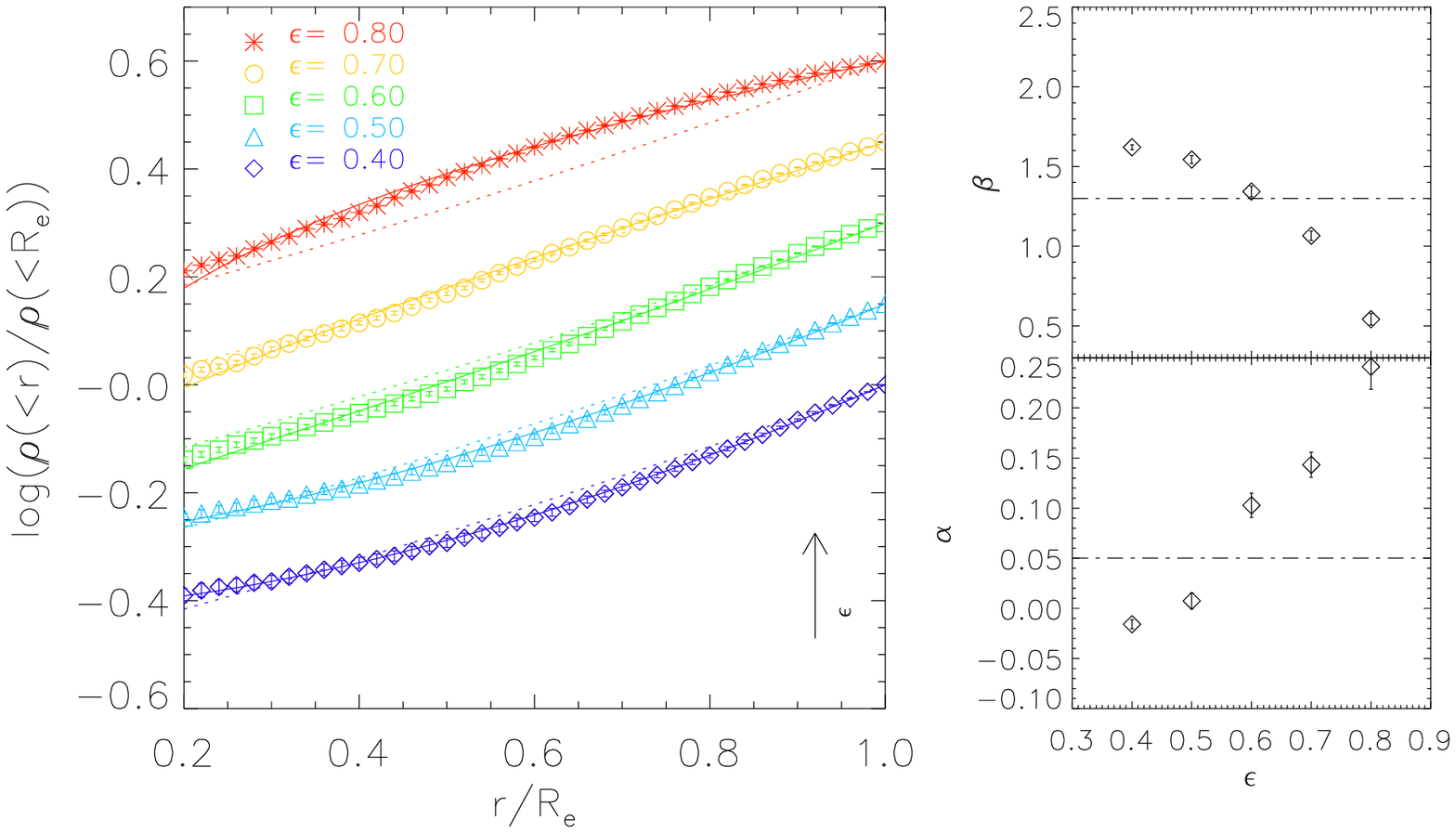}
\caption{The same as Figure \ref{profilesr} for voids stacked 
  according to their ellipticity. Equally spaced bins in
 ellipticity  have been used.}
\label{profilese}
\end{figure*}

We also explore the dependence of the void profiles on the global
properties of voids. In Figures \ref{profilesr}-\ref{profilese}, we show the profiles when
the void sample is divided according to: effective radius,
mean density contrast, inverse porosity and ellipticity. 
To enhance
the statistics in each subsample we have used all voids larger than
$3 \,h^{-1}\, Mpc$ in the last
five snapshots of the simulation (up to $z=0.25$).  As shown in Figure
\ref{profilesz}, the variation of the free parameters $\alpha$ and
$\beta$ in this redshift interval is quite modest,  hence we do not 
expect to introduce any bias in the analysis due to void evolution.

The trend with radius (Figure \ref{profilesr}) shows that the profiles are
stable and the same best-fit profile applies for voids with radius 
up to $R\sim 8
\,h^{-1}\, Mpc$. Indeed, these voids can be optimally described by the
same functional form used for the global population. 
However, there is an indication of steeper profiles in 
voids having higher radius (yellow and red curves). 
Nevertheless, as a consequence of the paucity of large voids  (see
Section \ref{stat}), the statistics in the sub-samples with the
highest radii is quite poor. Indeed, the mean profile results
significantly more noisy than that of small-sized voids.

When the void sample is splitted according to the mean density
contrast (Figure \ref{profilesd}), we observe a clear separation among
voids having different densities, with the least dense  voids
showing the steepest profile. Hence, void density appears as a much
more important parameter than void radius in determining the void structure.
The void profiles also show a systematic
dependence on void morphology.  Voids with high porosity (i.e. low
inverse porosity, blue curve
in Figure \ref{profilesi}) and high ellipticity (red curve in Figure
\ref{profilese}) show a sort of \textquoteleft bumpy\textquoteright\, profile, i.e. an enhanced
density in the interiors, the
reason being the high level of contamination in these sub-samples
which leads to higher density in the interiors.

\section{Summary and Conclusions}\label{conclusions}

The simulation used in this work has been specifically designed to
follow the formation and evolution of voids. It is based on an AMR
scheme that, contrary to the common practice, 
refines the most
underdense regions in the simulated volume.
For the identification of voids, we have developed a 
numerical algorithm optimally suited to find voids in AMR
simulations. 
Taking advantage of the AMR scheme, we have applied the procedure on
different levels of refinements (which correspond to different
numerical resolutions), obtaining voids on different hierarchical
levels. 
The goodness of the void finder has been proven with the aid  of a set of
mock voids having realistic profiles, showing that the algorithm is
able to reproduce the input void distribution with good
accuracy.

With this void finder, in  a cosmological box of comoving side length
$100 \,Mpc/h$, we identify $\sim$ 600 voids at redshift $z=0$ with sizes up to $17 \,Mpc/h$ 
and with typical density contrast within the void region
$\delta=-0.8$. 
More than half of the volume is filled by voids,
with a filling fraction $FF=0.58$.  
 In agreement with previous findings, we find that voids 
show an irregular morphology, being quite elongated and highly porous.
It is remarkable that the mean ellipticity that we find ($\epsilon=0.55$) agrees very
well with that  measured in other works on void shape,  that use completely
different void finders
\citep{Shandarin06, Platen08, Bos12}. 
This is a further demonstration of the reliability of 
the code. 

Thanks to the special refinement scheme adopted, we are able to resolve
the rich substructure contained within void. 
We have shown that subvoids 
appear as a rescaled version of their parent voids,
having smaller sizes, lower densities and more regular morphology,
being more spherical and less porous.

The morphological properties of voids  show a significant
evolution with redshift, with a systematic flattening of voids as time
proceeds and a growing porosity.
It is already known that void ellipticity should grow with time as a
result of the influence of the tidal field of the large scale structure \citep{Lee09,
  Bos12}. Similarly, we interpret the growing porosity as due to the void
surroundings. Indeed, if the voids  were evolving in  isolation, their porosity  could not
increase under  the action of  expansion only.  
Hence,  we are
prone to  ascribe the growing  porosity  to void  merging, that
is  the  inclusion  of  small  voids into  the  larger  ones
\citep{Vandew11}.

We have also analysed the mass density profile of voids larger than
$3\,Mpc/h$, rescaled to the density enclosed within the void effective
radius. 
To fit the void density profiles we propose a two-parameters
functional form, that is a generalization of the profile initially proposed by
\citet{Colberg05}.  We have shown that the same analytic expression can be used for voids at
different hierarchical levels, though voids in the bottom levels show 
flatter profiles, the reason being the absence of the sharp walls
present in the parent voids.
The evolution of the density profile with redshift shows a progressive
steepening of the profiles as cosmic time proceeds, in agreement
with the expectations from the linear theory \citep{Sheth04} of an
evolution towards a \textquoteleft bucket-shaped\textquoteright\, radial profile, as a consequence of
matter evacuation.  

The same functional form can also describe successfully voids having
different properties. Voids of  different sizes have density profiles
of very similar shapes that
can be fitted with the same best fit parameters. 
Although there is some evidence that the very large voids have a 
steeper profile than the global void population, the paucity of such large voids
prevents us to draw robust conclusions in this respect. 
There is nevertheless a significant dependence of the profile
steepness from the mean void density, being the least dense voids 
those with the steepest profile. Finally, void morphology also affects
the profile, in the sense that the most porous and elongated voids
display a sort of \textquoteleft bumpy\textquoteright\, profile, likely due to contamination from
regions not belonging to the void itself.

 It is worth to note that the fact that the same functional form for the void density profile
 can be applied to voids of any size, density, morphology
and redshift suggests the existence of  a universal density profile
for cosmic voids, in analogy  to the universal density profile for dark 
matter haloes. 

We conclude by noticing that the particular refinement scheme adopted
in this simulation is not only crucial to resolve the intricate web of
void substructures, but also optimized to resolve the
galaxy populations contained within voids. Although the study of void
galaxies is beyond the scope of this work, we plan to address this
issue in a forthcoming work.

\section*{Acknowledgements} 
We would like to thank the referee for his/her suggestions that helped improving the
manuscript. 
This work was  supported by the Spanish Ministerio de Econom\'{\i}a y Competitividad 
(MINECO, grants   AYA2010-21322-C03-01) and    the   Generalitat   Valenciana   (grant
PROMETEO-2009-103).  SP also acknowledges a fellowship from the
European Commission's Framework Programme 7, through the Marie Curie
Initial Training Network CosmoComp (PITN-GA-2009-238356) as well as
from the PRIN-INAF09 project \textquotedblleft Towards an Italian Network for Computational Cosmology\textquotedblright.

\label{lastpage}

\begin{thebibliography}{}

\bibitem[\protect\citeauthoryear{Aragon-Calvo 
\& Szalay}{2013}]{AragonCalvo13} Aragon-Calvo M.~A., Szalay A.~S., 2013, MNRAS, 428, 3409 

\bibitem[\protect\citeauthoryear{Beers, Flynn, 
\& Gebhardt}{1990}]{Beers90} Beers T.~C., Flynn K., Gebhardt K., 1990, AJ, 100, 32 


\bibitem[\protect\citeauthoryear{Bertschinger}{1985}]{Bert85} 
Bertschinger E., 1985, ApJS, 58, 1 

\bibitem[\protect\citeauthoryear{Bond et al.}{1991}]{Bond91} 
Bond J.~R., Cole S., Efstathiou G., Kaiser N., 1991, ApJ, 379, 440 

\bibitem[\protect\citeauthoryear{Bos et al.}{2012}]{Bos12} 
Bos E.~G.~P., van de Weygaert R., Dolag K., Pettorino V., 2012, MNRAS, 426, 
440 

\bibitem[\protect\citeauthoryear{Ceccarelli et 
al.}{2006}]{Ceccarelli06} Ceccarelli L., Padilla N.~D., Valotto C., 
Lambas D.~G., 2006, MNRAS, 373, 1440 


\bibitem[\protect\citeauthoryear{Colberg et 
al.}{2008}]{Colberg08} Colberg J.~M., et al., 2008, MNRAS, 387, 
933 

\bibitem[\protect\citeauthoryear{Colberg et 
al.}{2005}]{Colberg05} Colberg J.~M., Sheth R.~K., Diaferio A., 
Gao L., Yoshida N., 2005, MNRAS, 360, 216 

\bibitem[\protect\citeauthoryear{Colless et 
al.}{2001}]{Colless01} Colless M., et al., 2001, MNRAS, 328, 1039 

\bibitem[\protect\citeauthoryear{Croton et al.}{2004}]{Croton04} 
Croton D.~J., et al., 2004, MNRAS, 352, 828 

\bibitem[\protect\citeauthoryear{Crain et al.}{2009}]{Crain09} 
Crain R.~A., et al., 2009, MNRAS, 399, 1773 

\bibitem[\protect\citeauthoryear{Dekel 
\& Rees}{1994}]{Dekel94} Dekel A., Rees M.~J., 1994, ApJ, 422, L1 

\bibitem[\protect\citeauthoryear{Eisenstein \& Hu}{1998}]
{EiHu98} Eisenstein D.J., Hu W., 1998, ApJ, 511, 5

\bibitem[\protect\citeauthoryear{Gottl{\"o}ber et 
al.}{2003}]{Gottlober03} Gottl{\"o}ber S., {\L}okas E.~L., Klypin 
A., Hoffman Y., 2003, MNRAS, 344, 715 

\bibitem[\protect\citeauthoryear{Haart \& Madau}{1996}]
{hama96} Haart F.,  Madau P., 1996, ApJ, 461, 20

\bibitem[\protect\citeauthoryear{Hockney 
\& Eastwood}{1988}]{Hockney88} Hockney R.~W., Eastwood J.~W., 1988, csup.book,  

\bibitem[\protect\citeauthoryear{Hoyle et al.}{2012}]{Hoyle12} 
Hoyle B., Jimenez R., Verde L., Hotchkiss S., 2012, JCAP, 2, 9 

\bibitem[\protect\citeauthoryear{Hoyle 
\& Vogeley}{2004}]{HV04} Hoyle F., Vogeley M.~S., 2004, ApJ, 607, 751 

\bibitem[\protect\citeauthoryear{Icke}{1984}]{Icke84} Icke V., 
1984, MNRAS, 206, 1P 

\bibitem[\protect\citeauthoryear{Katz, Weinberg \& Hernquist}{1996}]
{katz96} Katz N., Weinberg D., Hernquist L., 1996, ApJS, 105, 19

\bibitem[\protect\citeauthoryear{Kirshner et 
al.}{1981}]{Kirshner81} Kirshner R.~P., Oemler A., Jr., Schechter 
P.~L., Shectman S.~A., 1981, ApJ, 248, L57 

\bibitem[\protect\citeauthoryear{Lavaux 
\& Wandelt}{2010}]{Lavaux10} Lavaux G., Wandelt B.~D., 2010, MNRAS, 403, 1392 

\bibitem[\protect\citeauthoryear{Lavaux 
\& Wandelt}{2012}]{Lavaux12} Lavaux G., Wandelt B.~D., 2012, ApJ, 754, 109 

\bibitem[\protect\citeauthoryear{Lee 
\& Park}{2009}]{Lee09} Lee J., Park D., 2009, ApJ, 696, L10 

\bibitem[\protect\citeauthoryear{Navarro, Frenk, 
\& White}{1997}]{NFW} Navarro J.~F., Frenk C.~S., White S.~D.~M., 1997, ApJ, 490, 493 

\bibitem[\protect\citeauthoryear{Neyrinck}{2008}]{Neyrinck08} 
Neyrinck M.~C., 2008, MNRAS, 386, 2101 

\bibitem[\protect\citeauthoryear{Neyrinck 
\& Yang}{2013}]{Neyrinck13} Neyrinck M.~C., Yang L.~F., 2013, arXiv, arXiv:1305.1629 

\bibitem[\protect\citeauthoryear{Padilla, Ceccarelli, 
\& Lambas}{2005}]{Padilla05} Padilla N.~D., Ceccarelli L., Lambas D.~G., 2005, MNRAS, 363, 977 

\bibitem[\protect\citeauthoryear{Pan et al.}{2012}]{Pan12} 
Pan D.~C., Vogeley M.~S., Hoyle F., Choi Y.-Y., Park C., 2012, MNRAS, 421, 
926 

\bibitem[\protect\citeauthoryear{Patiri et al.}{2006a}]{Patiri06a} 
Patiri S.~G., Betancort-Rijo J.~E., Prada F., Klypin A., Gottl{\"o}ber S., 
2006a, MNRAS, 369, 335 

\bibitem[\protect\citeauthoryear{Patiri et al.}{2006b}]{Patiri06b} 
Patiri S.~G., Prada F., Holtzman J., Klypin A., Betancort-Rijo J., 2006b, 
MNRAS, 372, 1710

\bibitem[\protect\citeauthoryear{Patiri, Betancort-Rijo, 
\& Prada}{2012}]{Patiri12} Patiri S.~G., Betancort-Rijo J., Prada F., 2012, A\&A, 541, L4 

\bibitem[\protect\citeauthoryear{Park 
\& Lee}{2007}]{Park07} Park D., Lee J., 2007, PhRvL, 98, 081301 

\bibitem[\protect\citeauthoryear{Platen, van de Weygaert, 
\& Jones}{2007}]{Platen07} Platen E., van de Weygaert R., Jones B.~J.~T., 2007, MNRAS, 380, 551 


\bibitem[\protect\citeauthoryear{Platen, van de Weygaert, 
\& Jones}{2008}]{Platen08} Platen E., van de Weygaert R., Jones B.~J.~T., 2008, MNRAS, 387, 128 

\bibitem[\protect\citeauthoryear{Plionis 
\& Basilakos}{2002}]{Plionis02} Plionis M., Basilakos S., 2002, MNRAS, 330, 399 

\bibitem[\protect\citeauthoryear{Quilis}{2004}]{quilis04}  Quilis  V.,
2004, MNRAS, 352, 1426

 \bibitem[\protect\citeauthoryear{Schaap 
\& van de Weygaert}{2000}]{Schaap00} Schaap W.~E., van de Weygaert R., 2000, A\&A, 363, L29 

\bibitem[\protect\citeauthoryear{Shandarin et 
al.}{2006}]{Shandarin06} Shandarin S., Feldman H.~A., Heitmann K., 
Habib S., 2006, MNRAS, 367, 1629 

\bibitem[\protect\citeauthoryear{Shectman et 
al.}{1996}]{Shectman96} Shectman S.~A., Landy S.~D., Oemler A., 
Tucker D.~L., Lin H., Kirshner R.~P., Schechter P.~L., 1996, ApJ, 470, 172 

\bibitem[\protect\citeauthoryear{Sheth 
\& van de Weygaert}{2004}]{Sheth04} Sheth R.~K., van de Weygaert R., 2004, MNRAS, 350, 517 

\bibitem[\protect\citeauthoryear{Springel \&
Hernquist}{2003}]{springe03} Springel  V., Hernquist L.,  2003, MNRAS,
339, 289

\bibitem[\protect\citeauthoryear {Sutherland \& Dopita}{1993}]
{sudo93} Sutherland R.,  Dopita M. S., 1993, ApJS, 88, 253

\bibitem[\protect\citeauthoryear{Tavasoli, Vasei, 
\& Mohayaee}{2013}]{Tavasoli12} Tavasoli S., Vasei K., Mohayaee R., 2013, A\&A, 553, A15 

\bibitem[\protect\citeauthoryear{Theuns et al.}{1998}]
{theuns98} Theuns T., Leonard A., Efstathiou G., Pearce F. R., Thomas P. A, 
1998, MNRAS, 301, 478

\bibitem[\protect\citeauthoryear{Tukey}{1958}]{Tukey58} Tukey J. W., 1958, Ann. Math. Stat, 29, 614

\bibitem[\protect\citeauthoryear{Yepes  et al.}{1997}]{yepes97} Yepes
G., Kates R., Khokhlov A., Klypin A., 1997, MNRAS, 284, 235

\bibitem[\protect\citeauthoryear{York et al.}{2000}]{York00} 
York D.~G., et al., 2000, AJ, 120, 1579 

\bibitem[\protect\citeauthoryear{van de Weygaert 
\& Platen}{2011}]{Vandew11} van de Weygaert R., Platen E., 2011, IJMPS, 1, 41 

\bibitem[\protect\citeauthoryear{van de Weygaert 
\& van Kampen}{1993}]{Vandew93} van de Weygaert R., van Kampen E., 1993, MNRAS, 263, 481 

\bibitem[\protect\citeauthoryear{Varela et al.}{2012}]{Varela12} 
Varela J., Betancort-Rijo J., Trujillo I., Ricciardelli E., 2012, ApJ, 744, 
82 

\bibitem[\protect\citeauthoryear{Vogeley et 
al.}{1994}]{Vogeley94} Vogeley M.~S., Geller M.~J., Park C., 
Huchra J.~P., 1994, AJ, 108, 745 

\end{thebibliography}
\end{document}